 \documentclass[10pt,preprint]{aastex}  %REQUIRED FOR APJ SUBMISSION
 \usepackage{natbib}
\def\gapprox{\lower.4ex\hbox{$\;\buildrel >\over{\scriptstyle\sim}\;$}}
\def\lapprox{\lower.4ex\hbox{$\;\buildrel <\over{\scriptstyle\sim}\;$}}

\shortauthors{ASCHWANDEN, KONTAR AND JEFFREY}
\shorttitle{Global energetics of Solar Flares VIII.}

\begin{document}

\title{Global Energetics of Solar Flares: VIII. The Low-Energy Cutoff}

\author{        Markus J. Aschwanden$^1$}
\affil{         $^1)$ Lockheed Martin,
                Solar and Astrophysics Laboratory,
                Org. A021S, Bldg.~252, 3251 Hanover St.,
                Palo Alto, CA 94304, USA;
                e-mail: aschwanden@lmsal.com }

%\and

\author{	Eduard P. Kontar$^2$ and Natasha L.S. Jeffrey$^3$}
\affil{		$^2)$ School of Physics and Astronomy,
		University of Glasgow,
		G12 8QQ Glasgow, UK;
		e-mail: natasha.jeffrey@glasgow.ac.uk}

\begin{abstract}
One of the key problems in solar flare physics is the determination of the low-energy cut-off; the value that determines the energy of nonthermal
electrons and hence flare energetics. We discuss different approaches
to determine the low-energy cut-off in the spectrum of accelerated electrons:
(i) the total electron number model,
(ii) the time-of-flight model
(based on the equivalence of the time-of-flight and the collisional
deflection time);
(iii) the warm target model of Kontar
et al.~(2015),
and (iv) the model of the spectral cross-over between
thermal and nonthermal components.
We find that the first three models are consistent with a low-energy cutoff
with a mean value of $\approx 10$ keV,
while the cross-over model provides an upper limit for the low-energy
cutoff with a mean value of $ \approx 21$ keV.
Combining the first three models we find that the ratio of the nonthermal
energy to the dissipated magnetic energy in solar flares has
a mean value of $q_E=0.57\pm0.08$, which is consistent with
an earlier study based on the simplified approximation of the
warm target model alone
($q_E=0.51\pm0.17$). This study corroborates the self-consistency
between three different low-energy cutoff models
in the calculation of nonthermal flare energies.
\end{abstract}


\section{	INTRODUCTION					}

The ultimate goal of this series of papers is the test of energy
closure in solar flares and associated coronal mass ejection (CME) events,
which entails the available energies that can be dissipated (magnetic
free energy $E_{mag}$, and aerodynamic drag energy $E_{drag}$),
and are transformed into primary energy dissipation
processes (acceleration of nonthermal particles $E_{nth}$,
direct heating $E_{dir}$, and the kinetic energy of a CME, $E_{cme}$),
as well as into secondary processes
(precipitation-induced thermal energies, and CME-accelerated particles).
Statistical results of these energies have been calculated for medium-sized
to large flare events (Emslie et al.~2012; Aschwanden et al.~2014,
2015; 2016; 2017; Aschwanden 2016, 2017; Aschwanden and Gopalswamy 2019).
A key result is the statistical energy closure of primary energy 
dissipation processes, i.e., 
$(E_{nth}+E_{dir}+E_{cme})/E_{diss}=0.87 \pm 0.18)$ (Aschwanden et al.~2017). 
The largest amount of the dissipated magnetic energy goes into the acceleration 
of electrons $E_{nth}/E_{diss}=0.51\pm0.17$. Importantly,
the measurement of the nonthermal energy $E_{nth}$ bears the largest
uncertainty due to the poorly known low-energy cutoff $\varepsilon_c$,
which is the central focus of this study.

The low-energy cutoff problem arises because the
instantaneous electron injection spectrum can be approximated with a
power-law function $f_e(\varepsilon) \propto \varepsilon^{-\delta}$ above
a minimum electron energy $\varepsilon_c$ (e.g., in the thick-target
model of Brown 1971). The fact that the power-law slope is generally
very steep, i.e., $\delta \approx 3-8$ (Dennis 1985), makes the
spectrally integrated electron flux extremely sensitive to the accurate
value of the low-energy cutoff value $\varepsilon_c$.
If we change this cutoff value from $\varepsilon=10$ keV by a factor
of 2 to $\varepsilon=20$ keV, the electron flux varies by a factor of
$\approx 2^{\delta}$, which amounts to 1-2 orders of magnitude.
The effects of low-energy cutoffs on solar flare microwave
and hard X-ray spectra was investigated in Holman (2003),
with the finding that microwave spectra become smoothed
in the optically thick portion, while hard X-ray (photon) spectra are
flattened below the cutoff energy. The modeling of the thermal
spectrum of hard X-ray photons has traditionally been done with an
isothermal model (Culhane et al.~1969; Culhane and Acton~1970;
Brown et al.~1974; Holman et al, 2011), while a multi-thermal
function involves a more realistic approach and was found to fit
the data equally well (e.g., Aschwanden 2007).
Moreover,  the altitude of the coronal X-ray sources are
observed to increase with energy in the thermal range (Jeffrey et al.~2015),
so that solar flares are multi-thermal and have strong vertical temperature 
and density gradients with a broad temperature distribution.
The ambiguity between an iso-thermal and a multi-thermal spectrum
contributes to further confusion between the thermal
and nonthermal spectral components, so that the spectral cross-over does
not reveal the exact cutoff energy, but yields a value that is about a factor of two too high. In a previous study on the multi-thermal modeling of 44 flare events,
the spectral cross-over was found in the range of $e_{co} = 10-28$ keV, with a
mean and standard deviation of $e_{co}=18.0 \pm 3.4$ keV (Aschwanden
2007).

A new theoretical model based on collisional relaxation and diffusion
of electrons in a warm coronal plasma was proposed by
Kontar et al.~(2015; 2019), which in principle yields the low-energy
cutoff $\varepsilon_{wt}$ in a modified thick-target model.
This modified thick-target model represents a more realistic approach,
because it generalizes the standard cold thick-target model (with a
cold plasma target) by including an additional warm plasma ``lid''
above the cold chromospheric component and, unlike the cold thick-target,
preserves the number of electrons in the warm plasma. Importantly,
the warm target model uses the warm coronal plasma environment (its
temperature, number density, and warm plasma extent) to constrain
the properties of the accelerated electron distribution. In general,
the low-energy cutoff should be determined by fitting the warm target
model to the observed X-ray count spectrum (see Kontar et al.~2019).
An application of a simplified version
of this warm target model to 191 M and X-class flares yielded
a mean low-energy cutoff of $\varepsilon_{wt}= 6.2\pm1.6$ keV
(Aschwanden et al.~2016), which is
significantly lower than the cross-over energy of $\varepsilon_{co}=21 \pm 6$ keV.
It can be shown that the low-energy cutoff in a cold thick-target model is
essentially undetermined (e.g. Ireland et al.~2013; Kontar et al.~2019),
while it was shown that the warm target model can constrain the
low-energy cutoff down to 7\% at a 3-$\sigma$ level (Kontar et al.~2019).

Here, we study the low-energy cutoffs inferred from the warm target model further. One issue is that the plasma in a flare is highly inhomogeneous,
ranging from the cold background corona values at the beginning
of a flare ($T_{cold}
\approx 0.5-2$ MK) to the hot chromospheric evaporation component
($T_{hot}\approx 5-25$ MK) at the flare peak time, causing some ambiguity about
which temperature to attribute to the warm-plasma component that
constrains the low-energy cutoff. In the warm target model, the deduction
of the coronal plasma environment is crucial for constraining the low
energy cutoff, and hence the nonthermal electron power (Kontar et al.~(2019).

Further, we will explore the total number of electrons in a flaring plasma and the spectral cross-over $\varepsilon_{co}$
as well as the warm target model $\varepsilon_{wt}$ predictions.
Moreover, the electron number model $\varepsilon_{en}$,
and the electron time-of-flight model $\varepsilon_{tof}$ will be applied.
The latter two models invoke the equivalence of the collisional
deflection time and the electron time-of-flight time scale,
as well as the limit of the maximum number of electrons that can be
accelerated in a finite flare volume, which at the same time
solves the electron number problem.

The content of this paper includes an analytical description
and derivation of all four theoretical models of the low-energy
cutoff (Section 2), followed by a description of the data analysis and
fitting of the theoretical models to the observational data sets
of all M and X-class flares observed with the {\sl Atmospheric Imaging
Assembly (AIA)} and the {\sl Helioseismic and Magnetic Imager (HMI)}
onboard the {\sl Solar Dynamics Observatory (SDO)} during
2010-2014, which amounts to 191 solar flare events (Section 3),
with discussion (Section 4) and conclusions (Section 5).

\section{	THEORY 						}

We describe four different models that independently provide
theoretical estimates of the low-energy cutoff of a hard X-ray
spectrum in solar flares.  In the following,
we present analytical derivations and assumptions of these models:
the {\sl electron number model} (Section 2.1), the {\sl time-of-flight
model} (Section 2.2), the {\sl warm target model} (Section 2.3),
and the {\sl spectral cross-over model} (Section 2.4).
The first two models are used here for the first time to derive
the low-energy cutoff, while the third model was used in Aschwanden
et al.~(2016), and the fourth model represents a common
method to derive upper limits on the low-energy cutoff.

\subsection{The Total Electron Number Model}

In the thick-target model (Brown 1971; e.g., see Section 13.2.2 in textbook Aschwanden 2004), the hard X-ray photon spectrum is defined
by a power law function of the observed photon energies $\epsilon_x$,,
\begin{equation}
	I(\epsilon_x) = I_1 {(\gamma-1) \over \epsilon_1}
	\left({\epsilon_x \over \epsilon_1} \right)^{-\gamma}
	\qquad [{\rm photons}\ {\rm cm}^{-2}\ {\rm s}^{-1}\
	{\rm keV}^{-1}] \ .
\end{equation}
The corresponding electron injection spectrum of electrons is,
\begin{equation}
	f_e(\varepsilon)
	= 2.68 \times 10^{33}\ (\gamma - 1)\ b(\gamma)\
	{I_1 \over \epsilon_1^2} \left({\varepsilon \over \epsilon_1}
	\right)^{-(\gamma + 1)}
	\qquad [{\rm electrons}\ {\rm keV}^{-1} {\rm s}^{-1}] \ ,
\end{equation}
and has the power law slope $\delta = \gamma+1$.
The total number of electrons above a cutoff energy $\varepsilon_c$, i.e.,
$F(\varepsilon \ge \varepsilon_c)$, is given by the thick-target model
\begin{equation}
	F(\varepsilon \ge \varepsilon_c)
	= \int_{{\varepsilon}_c}^\infty f_e(\varepsilon)\ d\varepsilon
	= 2.68 \times 10^{33}\ b(\gamma)\ {(\gamma-1) \over \gamma}
	{I_1 \over \epsilon_1} \left({\varepsilon \over \epsilon_1}
	\right)^{-\gamma}
	\qquad [{\rm electrons}\ {\rm s}^{-1}] \ ,
\end{equation}
where $b(\gamma)$ is an auxiliary function that contains the
beta function $B(p,q)$,
\begin{equation}
	b(\gamma) = \gamma^2 (\gamma-1)^2 B \left( \gamma - {1 \over 2},
	{3 \over 2} \right) \ ,
\end{equation}
which was been calculated by Hudson et al.~(1978) for a relevant range of
spectral slopes $\gamma$ of the observed photon spectrum, and
was approximated by the function (Aschwanden 2004),
\begin{equation}
	b(\gamma) \approx 0.27\ \gamma^3 \ ,
\end{equation}
and $\epsilon_1$ is the reference
energy at which the photon flux $I_1$ is measured.

Now we define the total number of electrons integrated over the
total flare duration $\tau_{flare}$,
\begin{equation}
	N_e = F(\varepsilon \ge \varepsilon_c) \ \tau_{flare}
	\qquad [{\rm electrons}] \ .
\end{equation}
On the other side, we can assume the total number of accelerated
nonthermal electrons during a flare by integrating the
preflare electron density $n_{e0}$ over the flare volume $V = L^3 q_{geo}$,
where $L$ is an appropriate length scale of a cube that encompasses
the entire flare volume,
\begin{equation}
	N_e = n_{e0} V
	= n_{e0}\ L^3\ q_{geo}
	\qquad [{\rm electrons}] \ ,
\end{equation}
and $q_{geo}$ is a geometric filling factor of the subvolume that
contains the number of electrons that can be accelerated out of the
cubic flare volume. We note that this assumption neglects the role of return
currents, which will maintain the total number of electrons (e.g., Somov 2000).
In other words, the total number of electrons in the flaring region is assumed
to be equal to the total number of electrons accelerated above the low energy
cut-off.  Even if this approximation is coarse, it gives useful details about
the efficiency of electron acceleration in solar magnetic reconnection regions.

In the standard CHSKP flare models for magnetic
reconnection (Carmichael 1964; Hirayama 1974; Sturrock 1966;
Kopp and Pneuman 1976), the subvolume
in which charged particles (electron and ions)
are accelerated encompasses about a fraction of $q_{geo} \approx 1/4$
of the cubic flare volume, as it can be estimated from the geometry shown in
Fig.~1 (shaded triangular subvolume).
The geometric filling factor consists of a factor of $q_{height}=1/2$
due to the vertical cusp range that covers half of the apex height,
and an additional factor of $q_{triangle}=1/2$ that accommodates
the ratio of the triangular arcade cross-section to the encompassing
cube volume, resulting into a combined factor of
$q_{geo} =q_{height} \times q_{triangle} = (1/2) \times (1/2) = 1/4$.
Alternatively, we can estimate the geometric filling factor from
the approximate size of the diffusion region of the magnetic reconnection
volume, which occupies the half
apex height ($h=L/2$) and half of the horizontal footpoint separation
($w_L/2$), and this way produces the same geometric filling factor of
$q_{geo}=(h/L) \times (w/L) = (1/2) \times (1/2) = 1/4$
(hatched area in Fig.~1).

Combining the two expressions for the total number of electrons
$N_e$ accelerated in a flare (using Eqs.~1-6) we obtain,
\begin{equation}
	N_e = {n_e L^3 q_{geo} }
	=  0.72 \times 10^{33}\ \gamma^2 (\gamma-1)
        {I_1 \over \epsilon_1} \left({\varepsilon_{en} \over \epsilon_1}
        \right)^{-\gamma}\ \tau_{flare}
        \qquad [{\rm electrons}] \ .	
\end{equation}
Using the normalized unit $L_{10}=L/10^{10}$ cm, we obtain the following
simple expression for the low-energy cutoff $\varepsilon_{en}$, where
the subscript "en" refers to the {\sl electron number} model,
\begin{equation}
	\varepsilon_{en} = \epsilon_1
	\left[ n_{e0}\ L_{10}^3\ q_{geo}\ \epsilon_1 \over
	0.72 \ \gamma^2 (\gamma-1)\ I_1\ \tau_{flare} \right]^{-1/\gamma}
	\qquad [{\rm keV}]\ ,
\end{equation}
which depends on the observables $n_{e0}, \gamma, I_1, \epsilon_1,
\tau_{flare}$ and
the model parameter $q_{geo} \approx 1/4$. The photon flux $I_1$ and
the spectral power law slope $\gamma$ at the energy $\epsilon_1$
can directly be obtained from a hard X-ray spectrum, the flare
duration $\tau_{flare}$ can be measured from hard X-ray time profiles,
and the electron density $n_{e0}$ has to be estimated before the onset
of the flare, which is typically $n_{e0} \approx 10^9$ cm$^{-3}$ (Fig.~3h).

Once we have a model for the low-energy cutoff $\varepsilon_{en}$,
we can calculate the power in nonthermal electrons above this
cutoff energy by integrating the electron energies $\varepsilon$,
with $b(\gamma)$ defined in Eqs.~(4) or (5),
\begin{equation}
	P_{en} (\varepsilon \ge \varepsilon_{en}) =
	\int_{\varepsilon_{en}}^\infty
	f_e(\varepsilon)\ \varepsilon\ d\varepsilon =
	4.3 \times 10^{24}\ b(\gamma) I_1
	\left({\varepsilon_{en} \over \epsilon_1}\right)^{-(\gamma-1)}
	\qquad [{\rm erg}\ {\rm s}^{-1} ] \
\end{equation}
and the total energy $E_{en}$ integrated over the flare duration
$\tau_{flare}=(t_2-t_1)$ is
\begin{equation}
	E_{en} = \int_{t_1}^{t_2} P_{en}(\varepsilon \ge \varepsilon_{en}, t)\ dt
	\qquad [{\rm erg}] \ ,
\end{equation}
where the photon flux $I_1(t)$, the power law slope $\gamma(t)$,
and the low-energy cutoff energy $\varepsilon_{en}(t)$ are
time-dependent.

\subsection{The Time-of-Flight Model}

For stochastic acceleration models with binary Coulomb collisions,
where particle gain and lose energy randomly,
the collisional mean free path yields an upper limit for the propagation
distance of free-streaming electrons.
The balance between acceleration and collisions can lead
to the formation of a kappa-distribution according to some solar flare
models (e.g., Bian et al.~2014). For solar flares, we can thus
estimate the critical energy between collisional and collisionless
electrons from the collisional deflection time $t_{defl}$
(Benz 1993),
\begin{equation}
	t_{defl} \approx 0.95 \times 10^8
	\left( {e_{keV}^{3/2} \over n_e} \right)
	\left( {20 \over \ln \Lambda} \right) \ ,
\end{equation}
where $\ln \Lambda \approx 20$ is the Coulomb logarithm. We set
the collisional deflection time equal
to the (relativistic) time-of-flight propagation time between
the coronal acceleration site and the chromospheric thick-target
energy loss site,
\begin{equation}
	t_{tof} = {L_{tof} \over v} = {L_{tof} \over \beta c} \ .
\end{equation}
The relativistic speed $\beta = v/c$,
\begin{equation}
	\beta = \sqrt{ 1 - {1 \over \gamma_r^2} } \ ,
\end{equation}
is related to the kinetic energy $e_{kin}$ of the
electron by,
\begin{equation}
	e_{kin} = m_e c^2 (\gamma_r - 1)
	=511 (\gamma_r - 1)\qquad {\rm [keV]} \ ,
\end{equation}
where $\gamma_r$ represents here the relativistic Lorentz factor
(not to be confused with the spectral slope $\gamma$ used above,
i.e., Eq.~(1)). We are setting these two time scales equal
(Aschwanden et al.~2016, Appendix A therein),
\begin{equation}
	t_{defl} = t_{tof} \ ,
\end{equation}
we use $\ln \Lambda \approx 20$, we define the kinetic energy
$\epsilon_{keV}=\epsilon_{kin}$, and obtain with Eqs.~(12-16),
\begin{equation}
	(\gamma_r - 1)^{3/2} \left( 1 - {1 \over \gamma_r^2} \right)^{1/2}
	= {L_{tof} \ n_e \over 0.95 \times 10^8 \times 511^{3/2} c}
	\ .
\end{equation}
Using the low-relativistic approximation (for $\gamma_r \gapprox 1$),
\begin{equation}
	(\gamma_r - 1)^{3/2} \left({1 - {1 \over \gamma_r^2}} \right)^{1/2}
	= (\gamma_r - 1)^{3/2} {(\gamma_r - 1)^{1/2} (\gamma_r + 1)^{1/2}
	\over \gamma_r} = {(\gamma_r - 1)^{2} (\gamma_r + 1)^{1/2} \over \gamma_r}
	\approx (\gamma_r - 1)^2 \sqrt{2} \ ,
\end{equation}
we obtain,
\begin{equation}
	(\gamma_r - 1)^2 \sqrt{2} \approx 0.0003 \times
	\left( {L_{tof} \over 10^{10}\ {\rm cm}} \right)	
	\left( {n_e \over 10^{10}\ {\rm cm}^{-3}} \right)	
	\ .
\end{equation}
The time-of-flight distance is approximately $L_{tof} = L \sqrt{2}$,
where the flare length scale $L$ is also the vertical extent of the cusp
(Fig.~1), and the factor $\sqrt{2}$ corrects for the mean pitch angle
($45^\circ$) of the electrons spiraling along the time-of-flight path.
Then, by inserting $(\gamma_r - 1) = e_c/$511 keV from Eq.~(15),
we find the cross-over energy $e_{tof} \approx e_{kin}$, explicitly
expressed,
\begin{equation}
	e_{tof} \approx 28
	\left( {L \over 10^{10}\ {\rm cm}} \right)^{1/2}	
	\left( {n_e \over 10^{10}\ {\rm cm}^{-3}} \right)^{1/2}	
	\qquad {\rm [keV]}\ .
\end{equation}
This expression requires the measurement of a mean length scale
$L=A^{1/2}$ of the flare area and an average electron density $n_e$
where flare-accelerated electrons propagate.

From the model of the low-energy cutoff energy $\varepsilon_{tof}$,
we can calculate the power in nonthermal electrons above this
cutoff energy by integrating over the electron energies $\varepsilon$,
\begin{equation}
	P_{tof} (\varepsilon \ge \varepsilon_{tof}) =
	\int_{\varepsilon_{tof}}^{\infty} f(\varepsilon) \ d\varepsilon =
	4.3 \times 10^{24}\ b(\gamma) I_1
	\left({\varepsilon_{tof} \over \epsilon_1}\right)^{-(\gamma-1)}
	\qquad [{\rm erg}\ {\rm s}^{-1} ] \ . 
\end{equation}
The total energy integrated over the flare duration is then, using the
time-dependent functions $\gamma(t)$, $I_1(t)$, and $\varepsilon_{tof}(t)$,
\begin{equation}
	E_{tof} = \int_{t_1}^{t_2} P_{tof}
	(\varepsilon \ge \varepsilon_{tof}, t)\ dt
	\qquad [{\rm erg}] \ .
\end{equation}

Turning the argument around predicts a time-of-flight distance
$L_{tof} \approx \varepsilon_{tof}^2/n_e$ as a function of the low-energy cutoff
$\varepsilon_{tof}$, which is a similar concept that has been applied to model
the size $L$ of the acceleration region as a function of the
electron energy $e$, i.e., $(L-L_0) \propto e^2/n_e$
(Guo et al.~2012a, 2012b, 2013; Xu et al.~2008).

\subsection{The Warm-Target Model}

Previous applications of the thick-target model
generally assume cold (chromospheric) temperatures in the
electron precipitation site (e.g., Holman et al.~2011, for a
review). At the same time, the temperature of the flaring solar
corona is sufficiently high so that finite temperature effects must be included
(Galloway et al.~2005; Goncharov et al.~2010; Jeffrey et al.~2014).
Moreover, the slow spatial diffusion of thermalized electrons, previously ignored, led to the theoretical development of the warm target model (Kontar et al.~2015).
The model has been tested with numerical simulations that include the effects
of collisional energy diffusion, spatial transport and thermalization of fast
electrons (Jeffrey et al.~2014).

The warm target model assumes a two-temperature target plasma
(Kontar et al.~2015, 2019): the warm solar corona and
the cold chromosphere. The warm corona is collisionally thick
to electrons with energy $E < \sqrt{2 K n L}$, where
$K = 2 \pi e^4 \ln \Lambda$ is a constant, $n$ is the density
of the coronal plasma, and $L$ is the length of the warm target
region. Therefore, the accelerated electrons injected into a
flaring loop propagate and collide in the warm plasma. Electrons
with energy $E^2 < 2 K n L$ lose all of their energy in the
coronal plasma and join the Maxwellian distribution of the
surrounding plasma, increasing the density of thermal plasma
in the loop. The mean electron flux spectrum can be represented
by (Kontar et al.~2015),
\begin{equation}
	\langle n V F \rangle (E) = {1 \over 2 K} E e^{-E/k_BT}
	\int_{E_{min}}^E
	{e^{E^\prime/kT} dE^\prime \over E^\prime G
	\left(\sqrt{E^\prime \over k_B T}\right)}
	\int_{E^\prime}^{\infty} \dot{N}(E_0) dE_0 \ ,
\end{equation}
where $G(x)=[{\rm{erf}(x)} - {x}\ \rm{erf}^\prime{ (x)}]/2 { x}^2$.
The lower limit in Eq.~(23) is given by,
\begin{equation}
	E_{min} \approx 3 k_B T \left({5 \lambda} \over L \right)^4 \ ,
\end{equation}
where $\lambda = (k_B T)^2/2 K n$ is the collisional mean free path,
and Eq.~(24) is determined by considering the warm plasma properties
in the corona. The mean electron flux $\langle n V F\rangle(E)$
convolved with the bremsstrahlung cross-section $ \sigma(E, \epsilon)$
predicts the X-ray flux spectrum at $R=1$ AU, 
\begin{equation}
 	I(\epsilon)=\frac{1}{4\pi R^2}\int_{E}^{\infty}
	\langle n V F \rangle (E) \sigma(E, \epsilon) dE \ .
\end{equation}
where $\epsilon$ is the photon energy. 
Fitting the warm target model X-ray spectrum to the observed X-ray spectrum,
allows us to determine the parameters of the injected electron flux spectrum,
which here is assumed to be a power-law\footnote{A warm target kappa model
is also available in OSPEX (see Kontar et al.~(2019)).}
\begin{equation}
	\dot{N}(E)=\dot{N}_{0} \frac{\delta-1}{E_{c}}
	\left(\frac{E}{E_{c}}\right)^{-\delta} \ ,
\end{equation}
where $\dot{N}_{0}$ is the electron acceleration rate [electrons/s],
$\delta$ is the spectral index, and $E_c$ is the low-energy cut-off
in the injected electron spectrum.

The warm target model suggests that electrons are thermalized in the
warm plasma of the coronal loop and produce detectable thermal emission
with an emission measure of,
\begin{equation}
	\Delta EM \approx {\pi \over K} \sqrt{{m_e \over 8}}
	(k_B T)^2 {\dot{N_0} \over {E_{min}^{1/2}}} \ ,
\end{equation}
where $\Delta EM$ characterizes the additional contribution to the
soft X-ray emission measure from the thermalized accelerated electrons.
These hot Maxwellian electrons can diffusively escape from the warm
plasma of the loop and collisionally stop in the dense cold
chromosphere. High-energy electrons with $E^2 > 2 K n L$ behave
in the same way as in the standard cold thick-target model.
It is important to note that the warm target model is responsible
for the nonthermal component, and for a fraction of the thermal component
of the X-ray emission.
The pile-up of low energy electrons thermalized in the flaring corona
allows us to solve the low energy cut-off problem (Kontar et al.~2019)
by comparing the thermalized electrons,
that is, by determining the contribution from Eq.~(27) and the observed
X-ray spectrum.
In other words, if the low-energy cutoff is determined too low
(i.e. if the contribution from $\Delta$ EM is too large),
then the warm target model produces too many thermalized electrons
and hence can be ruled out.

According to the warm-target model of Kontar et al.~(2015),
the effective low-energy cutoff $E_c\simeq \varepsilon_{wt}$ can be
coarsely approximated as
\begin{equation}
	\varepsilon_{wt} \approx (\xi + 2) k_B T_e = \delta \ k_B T_e \ ,
\end{equation}
where $\xi = \gamma - 1$ is the power-law slope of the
source-integrated mean electron flux spectrum (see Eqs.~(8)-(10)
in Kontar et al.~2015), and $T_e$ is the temperature of the
warm target plasma. For medium-sized to large X-class flares, this temperature range spans
$T_e \approx 10-30$ MK, giving (in energy units)
$e_{th} = k_B T_e = 0.9-2.6$ keV, and for a typical value
of the photon spectral slope $\delta = \gamma + 1 \approx 4$,
low-energy cutoffs of $e_{th} = \delta \ k_B T_e \approx
3.5-8.5$ keV are predicted. In this simplified version,
Kontar et al.~(2015) stress that the value of $T_e$ used must be
the value of $T_e$ corresponding to the Maxwellian thermal plasma in the loop.

Further, we stress that Eq.~(28) is determined by considering the energy at
which the systematic energy loss rate vanishes in the Fokker-Planck equation
governing the evolution of $\langle n V F \rangle$ in a warm plasma, 
and that an accurate
determination of the properties of the accelerated electron distribution can
only be determined using the combination of X-ray spectroscopy and imaging
outlined in detail in Kontar et al.~(2019). We note, that while the expression
(Eq.~28) is an approximation only, it does allow for a relatively robust
statistical analysis (Aschwanden et al 2017), while the detailed
fitting outlined in Kontar et al.~(2019) is challenging for a large number
of flare events. However, the detailed fitting procedure of
Kontar et al.~(2019), which constrains the plasma parameters $T_e$, $n$ and $L$,
is the recommended way to determine the nonthermal
electron properties in an individual flare. 
Here, the use of Eq.~(28) is likely to provide a lower limit of $e_{th}$, 
but is still useful for the purpose of a large statistical study.

From the low-energy cutoff approximation $\varepsilon_{wt}$,
we can calculate the power in the electron flux $P_{wt}$,
\begin{equation}
	P_{wt} (\varepsilon \ge \varepsilon_{wt}) =
	\int_{\varepsilon_{wt}}^{\infty} f(\varepsilon) \ d\varepsilon =
	4.3 \times 10^{24}\ b(\gamma) I_1
	\left({\varepsilon_{wt} \over \epsilon_1}\right)^{-(\gamma-1)}
	\qquad [{\rm erg}\ {\rm s}^{-1} ] \
\end{equation}
and the total energy integrated over the flare duration is
\begin{equation}
	E_{wt} = \int_{t_1}^{t_2} P_{wt}
	(\varepsilon \ge \varepsilon_{wt}, t)\ dt
\end{equation}

\subsection{The Spectral Cross-Over Model}

The bremsstrahlung spectrum $I(\varepsilon)$ of a thermal plasma
with electron temperature $T_e$, as a function of the photon
energy $\varepsilon = h \nu$ (with $h$ the Planck constant
and $\nu$ the frequency), setting the coronal electron density
equal to the ion density ($n = n_i = n_e$), and neglecting
factors of the order of unity (such as the Gaunt factor
$g(\nu, T)$ in the approximation of the Bethe-Heitler
bremsstrahlung cross-section), and the ion charge number,
$Z \approx 1$, is (Brown 1974; Dulk and Dennis 1982),
\begin{equation}
	I(\varepsilon) = I_0 \int
	{ \exp{( - \varepsilon/k_B T )} \over T^{1/2}}
	{dEM(T) \over dT} \ dT \ ,
\end{equation}
where $I_0 \approx 8.1 \times 10^{-39}$ keV cm$^{-2}$ s$^{\-1}$
keV$^{-1}$ and $dEM(T)/dT$ specifies the differential emission
measure (DEM) $n^2 dV$ in the volume $dV$
corresponding to a temperature range of $dT$,
\begin{equation}
	\left( {dEM(T) \over dT} \right) \ dT
	= n^2(T) \ dV \ .
\end{equation}
Regardless, whether we define this DEM distribution by an
isothermal or by a multi-thermal plasma (Aschwanden 2007),
the thermal spectrum $I(\varepsilon)$ falls off similarly
to an exponential function at an energy of $\varepsilon
\lapprox 20$ keV (or up to $\lapprox 40$) keV in extremal
cases), while the nonthermal spectrum in the higher
energy range of $\varepsilon \approx 20-100$ keV can be
approximated by a single (or broken) power-law function (Eq.~3).

Because of the two different functional shapes, a cross-over
energy $\varepsilon_{co}$ can be defined by the change
in the spectral slope between the thermal and the nonthermal
spectral component. The electron energy spectrum, however,
can have a substantially lower or higher cutoff energy (e.g.,
Holman 2003). We represent the combined spectrum with the
sum of the (exponential-like) thermal and the (power-law-like)
nonthermal component, i,e.,
\begin{equation}
	I(\varepsilon) = I_{th}(\varepsilon) + I_{nth}(\varepsilon)
	= I_0 \int {\exp{( - \varepsilon/k_B T )} \over T^{1/2}}
	{dEM(T) \over dT} \
        + I_1 {(\gamma-1) \over \epsilon_1}
        \left({\epsilon_x \over \epsilon_1} \right)^{-\gamma} \ ,
\end{equation}
where the cross-over energy $\varepsilon_{co}$ can be determined in the
(best-fit) model spectrum $I(\varepsilon)$ from the energy where
the logarithmic slope is steepest, i.e., from the maximum of
$\partial \log I(\varepsilon)/\partial \log \varepsilon$.
The change of the spectral slope between the thermal and the
nonthermal component is depicted in Fig.~2, where cross-over
energies of $\varepsilon_{co}=4.7$ keV for a small flare is
calculated, and $\varepsilon_{co}=19.5$ keV for a large flare.

From the low-energy cutoff $\varepsilon_{co}$ we can calculate
the power in the electron flux $P_{co}$,
\begin{equation}
	P_{co} (\varepsilon \ge \varepsilon_{co}) =
	\int_{\varepsilon_{co}}^{\infty} f(\varepsilon) \ d\varepsilon =
	4.3 \times 10^{24}\ b(\gamma) I_1
	\left({\varepsilon_{co} \over \epsilon_1}\right)^{-(\gamma-1)}
	\qquad [{\rm erg}\ {\rm s}^{-1} ] \ ,
\end{equation}
and the total energy integrated over the flare duration is
\begin{equation}
	E_{co} = \int_{t_1}^{t_2} P_{co}
	(\varepsilon \ge \varepsilon_{co}, t)\ dt \ .
\end{equation}
We should be aware that the so determined cross-over energy
$\varepsilon_{co}$ is an upper limit only, and consequently
the total energy $E_{co}$ is a lower limit, unlike the
other three low-energy cutoff models described in Sections 2.1-2.3.

\section{	OBSERVATIONS AND DATA ANALYSIS			}

The previously analyzed data set is based on all M and X-class
flares observed with the {\sl Atmospheric Imaging
Assembly (AIA)} (Lemen et al.~2012) and the 
{\sl Helioseismic and Magnetic Imager (HMI)} (Scherrer et al.~2012)
onboard the {\sl Solar Dynamics Observatory (SDO)} 
spacecraft (Pesnell et al.~2011) during 2010-2014,
which amounts to 399 solar flare events. Here we use only those events
that have been simultaneously observed with the {\sl Ramaty
High Energy Solar Spectroscopic Imager (RHESSI)}
(Lin et al.~2002), which amount
to 191 events, due to the duty cycle of $\approx 50\%$ of RHESSI 
when the orbit is on the sunward side.

\subsection{	Spectral Modeling of RHESSI Data		}

We use the same RHESSI data of 191 flare events as previously
analyzed in Aschwanden et al.~(2016), using the OSPEX
(Object Spectral Executive) software (http://hesperia.gsfc.nasa.gov/).
We re-analyzed the RHESSI data by optimizing the flare time intervals
and the energy intervals (typically in the fitting range of $\varepsilon
\approx 10-30$ keV) and obtained essentially the same results as
described in Aschwanden et al.~(2016). The observed hard X-ray
photon spectrum has been fitted with an isothermal component
(that is defined by the emission measure $EM_{49}$ in units
of $10^{49}$ [cm$^{-3}$] and the temperature $T_e$ in units
of [MK]), plus a nonthermal component with a broken power law function
(that is defined by the nonthermal flux $I_{1}$ in units of
[photons cm$^{-2}$ s$^{-1}$ keV$^{-1}$] at a reference energy of
$\epsilon_1=50$ keV, and by the power law index $\delta$ of the
fitted (lower) electron spectrum, which corresponds to a power law index
of $\gamma = \delta - 1$ in the thick-target model. Examples
of such two-component (thermal plus nonthermal) hard X-ray photon
spectra are illustrated in Fig.~2. The hard X-ray
spectra are fitted in time intervals of $\Delta t = 20$ s and
yield the time-dependent best-fit parameters $EM_{49}(t)$,
$T_e(t)$, $I_{nth}(t)$, and $\delta(t)$. The maximum values
of the emission measure $EM_{49}$, the temperature $T_e^{rhessi}$, and
the photon flux $I_1(t)$, during the flare duration $\tau_{flare}$,
as well as the minimum value of the spectral slope
$\gamma=\delta-1$, are listed in Table 2 for 160 (out of the 191)
available events (omitting the less reliable cases with data gaps
or inaccurate fits that result into outliers with extreme
nonthermal energies of $E_{nth} > 10^{33}$ erg).
More details of the spectral modeling of RHESSI data are given
in Section 3.1 in Aschwanden et al.~(2016).

\subsection{	Differential Emission Measure (DEM) Modeling    }

Besides the hard X-ray spectral modeling, we need also to measure
the parameters of the spatial length scale $L$, the electron
temperature $T_e$ and the electron density $n_e$ during the
preflare phase as well as during the flare.
The preflare electron density $n_{e0}$ and the mean flare electron
density $n_e$ are listed in the three last columns of Table 2,
i.e., labeled as $b_{10}=n_e^{bg}/10^{10}$ for the background
and $n_{10}=n_e^{flare}/10^{10}$ during the flare.

The spatial length scale $L$ has been deduced from measuring
the flare area $A(t) = L(t)^2$, subject to corrections due
to projection effects and electron density scale heights
$\lambda$ (Aschwanden et al.~2014, 2015), where the flare
volume $V$ is approximated by the Euclidean relationship,
\begin{equation}
	V = L^3 \ .
\end{equation}

From {\sl differential emission measure (DEM)} modeling of the
EUV data (observed with AIA) earlier (Aschwanden et al.~2015), we obtained the
emission measure $EM_{EUV}$ of the (``cold'' and ``warm'') flare plasma
and emission-measure-weighted temperature ($T_{EUV})$,
and the corresponding electron density ($n_{EUV}$),
\begin{equation}
	n_{EUV} = \sqrt{EM_{EUV} \over V} \ ,
\end{equation}
measured at the peak time of the nonthermal hard X-ray flux.

In addition, the thermal emission measure ($EM_R$)
and temperature $T_R$ of the hot'' flare plasma
has been measured from the 2-component (thermal and
nonthermal) spectral fit to the RHESSI data, but we should
be aware that the RHESSI-inferred values are always biased
towares the hottest temperature component. Nevertheless, the
corresponding electron density $n_R$ is then
defined by the relationship during the flare at times $t$,
\begin{equation}
	n_R(t) = \sqrt{EM_R(t) \over V} \ .
\end{equation}
Measuring the density at the starting time of the flare ($t=t_1$)
yields then also an estimate of the preflare (or background) density
($n_{bg}$),
\begin{equation}
	n_{bg} = n_R(t=t_1) = \sqrt{EM_R(t_1) \over V} \ .
\end{equation}
This preflare density $n_{bg}$ is used in the electron number
model (Section 2.1), where the maximum possible number of
accelerated electrons in the full flare volume (essentially
defined by the envelope volume of the entire flare arcade)
during the preflare phase corresponds to the partial volume $V=L^3 q_{geo}$
(Eq.~6), with a geometric filling factor $q_{geo}=1/4$ derived
from the geometry of the diffusion region in a 3-D magnetic
reconnection process with propagation of the hard X-ray footpoints
along a flare ribbon with an approximative length $L$.

In the time-of-flight model (Section 2.2) we need an electron
density $n_e$ that is representative of the hot evaporating
plasma, where electrons are stopped by collisional deflection.
For this regime we use the emission measure $EM_R(t)$ and
temperature $T_R(t)$ that is obtained from the
spectral fitting of the thermal component observed with RHESSI.

In the warm target model (Section 2.3) we need an electron
temperature that is characteristic for the ``warm''
target region (from the acceleration region to the top of
the chromosphere), where the thermalization of fast electrons
takes place. 
We estimate this intermediate temperature
from the geometric mean of the ``warm'' plasma observed
in EUV (used in the DEM analysis) and the ``hot'' thermal
plasma seen by RHESSI,
\begin{equation}
	T_e(t_p) = \left[ T_{EUV} \times T_R(t_p) \right]^{1/2} \ .
\end{equation}
The temperature during the peak time $t_p$ of the nonthermal
hard X-ray flux is listed in Table 2, and a histogram is shown in Fig.~(3a),
which reveals a typical range of $T_e \approx 5-30$ MK.

\subsection{	Statistical Results				}

The statistical distributions of the observables are shown in
form of histograms on a linear or logarithmic scale in Fig.~3
and are listed in Table 1. The median values are:
$T_e \approx 12.5$ MK for the maximum electron temperature (defined by
the geometric mean between the EUV-inferred ($T_{EUV}$) and
RHESSI-inferred ($T_{R}$) values); $\gamma \approx 7$ for the photon spectral index;
$L \approx 10$ Mm for the spatial flare length scale; $\tau_{flare}
\approx 0.5$ hrs for the flare duration (defined by the time difference
between GOES start and peak times); $EM \approx 1 \times 10^{47}$ cm$^{-3}$
for the emission measure observed by RHESSI; $F \approx 5 \times 10^{-4}$
[photons cm$^{-2}$ s$^{-1}$ keV$^{-1}]$ for the photon flux
at $\epsilon_1 = 50$ keV; $n_{eo} \approx 1 \times 10^{10}$ cm$^{-3}$ for
the preflare electron density; and $n_e \approx 2 \times 10^{10}$
cm$^{-3}$ for the maximum flare electron density.

The statistical results of this analysis consist of the
low-energy cutoffs $\varepsilon_c$ and the total nonthermal energies
$E_{nth}$ of 191 M and X-class flares for all four theoretical
models, which are tabulated in Fig.~3, while the size
distribution of the low-energy cutoffs are displayed in Fig.~4,
and the size distributions of nonthermal energies are shown in
Fig.~5.

The size distributions of the low-energy cutoffs shown in Fig.~4
reveal almost identical median values for the first three models,
$\varepsilon_{en}=10.8$ keV for the electron number model (Fig.~4a),
$\varepsilon_{tof}=9.8$ keV for the time-of-flight model (Fig.~4b), and
$\varepsilon_{wt}= 9.1$ keV for the warm target model (Fig.~4c),
while the cross-over model reveals a value that is a factor of 2 higher,
i.e., $\varepsilon=21$ keV, which clearly corroborates the
theoretical expectation that the spectral cross-over represents
an upper limit on the low-energy cutoff only. Now we have a
quantitative result that the low-energy cutoff is over-estimated
by a factor of 2, statistically. This has the consequence that
the nonthermal energy is underestimated by about a factor of about
$2^4=16$ (for an electron power index of $\delta \approx 4$).

The size distributions of the nonthermal flare energies of the
analyzed 191 flare events are displayed in Fig.~5, for each of
the 4 low-energy cutoff models separately. The most conspicous
difference between the different theoretical models is that
the cross-over model is not able to produce nonthermal energies
above $E_{nth} \gapprox 2 \times 10^{30}$ erg, while the other three models
all can produce energies up to $E_{nth} \lapprox 10^{33}$ erg.
This is consistent with the expected bias that upper limits of
the low-energy cutoff substantially underestimate the spectral
integrated energy for the cross-over model,
because the nonthermal energy scales with a very high
nonlinear power (typically with a power index of $\delta \approx 4$).
There are additional differences in the size distributions,
especially regarding the power law index of the slope. The
electron number model produces a negative power law slope of
$\alpha \approx 1.4$, which is closest to most energy distributions
of solar flares among the first three models shown in Fig.~5
(e.g., $\alpha_E = 1.53$; Crosby et al.~2013). The warm target model
produces a surprisingly flat power law slope, with $\alpha \approx 1.1$,
probably because of a systematic overestimation of
the nonthermal energy of large flares. It is possible that
the functional form of the low-energy cutoff spectrum, for which
traditionally a step function at the lower boundary
$\varepsilon_c$ is assumed (e.g., Holman 2003), may be unrealistic.
A smoother function for the boundary would steepen the power law
slopes of the size distributions for the warm target model and
the time-of-flight model, and this way would bring them closer
to the canonical value of $\alpha_E \approx 1.5$ observed in
nonthermal energies (e.g., Crosby et al.~2013; see Table 3 in
Aschwanden 2015).

\subsection{	Nonthermal Energy versus Dissipated Magnetic Energy }

The main focus of this series of studies is the global energetics
and energy partition in solar flares and {\sl coronal mass
ejections (CMEs)}. One of the previous results is that the
nonthermal energy $E_{th}$ as a fraction of the dissipated
magnetic free energy $E_{diss}$ is $q_E=E_{nth}/E_{diss}
=0.51\pm0.17$, so about half of the dissipated magnetic
energy is converted into acceleration of electrons
(Aschwanden et al.~2017). If we
plot the same ratios for each of the theoretical models, we find
$q_E^{ne}=0.40\pm0.10$ for the electron number model (Fig.~6a),
$q_E^{wt}=0.45\pm0.10$ for the warm target model (Fig.~6b),
$q_E^{tof}=0.58\pm0.16$ for the time-of-flight model (Fig.~6c), and
$q_E^{co}=0.0034\pm0.0006$ for the cross-over model (Fig.~6d).

Since the three methods of calculating the nonthermal energy
are essentially independent, we can improve the accuracy of
the statistical means by averaging (logarithmically) the
values from two or three models, which is shown in Fig.~7.
Combining the electron number
and the warm target model, we find $q_E^{ne,wt}=0.57\pm0.10$ (Fig.~7a),
combining the electron number and the time-of-flight method
we find $q_E^{ne,tof}=0.52\pm0.09$ (Fig.~7b), or by combining the warm target
and the time-of-flight model we find $q_e^{wt,tof}=0.61\pm0.10$ (Fig.~7c).
The largest statistics is achieved by combining all three methods
(excluding the cross-over model), for which we find
$q_e^{en,wt,tof}=0.57\pm0.08$ (Fig.~7d), which is perfectly
consistent with the earlier result of $q_E=0.51\pm0.17$
(Aschwanden et al.~2017). However, the new result has a smaller
error of the mean ($q_{err}=\pm 0.07$) than the old result
($q_{err}=\pm 0.17$), thanks to the larger statistics with
multiple independent methods, which cancel out some of the
systematic errors of the various models. Note that the
uncertainty of the ratio of the nonthermal to the dissipated magnetic
energy, i.e., $q_E=E_{diss}/E_{magn}$, has been reduced to a factor of
$\sigma \approx 5$ for a single flare event (Fig.~7d), while the error of
the mean has been reduced to $q_{err}=0.08$ (Fig.~7d).

\section{	DISCUSSION					}

\subsection{	Constraints for Low-Energy Cutoffs		}

We applied four different theoretical considerations in order to determine
low-energy cutoffs in hard X-ray spectra, which are useful to pinpoint
systematic errors of the models. Let us discuss which parameters
constrain the various models, and whether the four models
have some common physics.

In the electron number model (Section 2.1) we make the assumption
that all electrons in the diffusion region of a magnetic reconnection
volume are accelerated out of the thermal distribution,
and therefore the flare volume $V=L^3$, the preflare electron
density $n_e$, and the flare duration $\tau_{flare}$ are needed,
as well as the observables that characterize the nonthermal spectrum
($I_1, \epsilon_1, \gamma$). This method, therefore requires
imaging observations (to measure the flare area $A=L^2$) and
time profiles of the photon flux $F(t)$ (to measure the flare duration),
while less physical parameters are required in the other models, and
thus the electron number model provides the strongest constraints
on the low-energy cutoff.

In the time-of-flight model (Section 2.2) we assume the
equivalence between collisional deflection and electron time-of-flight
times, which depend on the kinetic energy of electrons and the
electron density, plus the spatial scale of the electron time-of-flight
distance $L_{tof}$. Hence imaging observations are required also,
but the low-energy cutoff depends on $L_{tof}$ and $n_e$ only,
which amounts to less constraints than the electron number model.

In the simplified approximation of the  
warm target model (Section 2.3), only the temperature $T_e$ is
required to characterize the collisional loss in the thick-target
model (besides the spectral observable $\gamma$),
which is based on the same physical process of collisional thermalization
as the time-of-flight model, but requires less physical parameters.

Finally, in the spectral cross-over model (Section 2.4), the low-energy
cutoff is directly estimated from the cross-over of the thermal and
nonthermal spectrum, which does not require the knowledge of any
physical parameter. However, this simplest method provides
upper limits on the low-energy cutoff only.

So, the four methods are all complementary and at this point we
cannot claim which model has a systematically higher value for
the calculation of the low-energy cutoff, except for the fourth
method that provides upper limits on the low-energy cutoff only.
How compatible are the different models ?
For the scaling of the physical parameters $L$ and $n_e$
in the two first models, we find $\varepsilon_{en} \propto
 (n_e L^3)^{-1/\gamma}$ for the electron number model (Eq.~9), and
$\varepsilon_{tof} \propto (n_e L)^{1/2}$ for the time-of-flight
model (Eq.~20), which is not directly compatible, and thus indicates
incomplete physical models.

\subsection{	Functional Shape of Low-Energy Cutoff 		}

In most previous, work the functional shape of the (nonthermal)
electron injection spectrum is characterized with a power law function,
i.e., $f(\varepsilon \ge \varepsilon_{c}) \propto \varepsilon^{-\delta}$,
with a sharp cutoff at the low-energy side of the spectrum.
This functional choice of the spectrum is not constrained by any physical model,
but is simply chosen for mathematical convenience.
The steep fall-off of this function at $\varepsilon \ge \varepsilon_c$ creates
a particle energy distribution peaking near $\varepsilon_c$, 
which is unlikely to occur in a collisional plasma. 
We can use a kappa-distribution instead, already implemented in OSPEX.
There are very few studies of the low-energy cutoff with
smooth functions, such as modeling with kappa-distributions
(Bian et al.~2014; Kontar et al.~2019).

Alternatively, we derived 
a smooth low-energy cutoff function in Appendix A, which is not based
on a physical model either, but represents the simplest spectral
function with a low-energy cutoff at the lower end and a power-law
function at the upper end (Eq.~A1).
We show an example in Fig.~8, where the smooth low-energy cutoff
function (according to Eq.~A1) is shown with a minimum energy of
$\varepsilon_c=10.0$ keV, a power law slope of $\delta=3$,
and a peak energy of $\varepsilon_{peak}=\varepsilon_{min}
(1 + 1/\delta)=13.3$ keV.
Although the difference of the sharp and the smooth electron injection 
spectrum does not appear to be paramount on a log-log scale (Fig.~8, left), 
the same functions rendered on a linear scale (Fig.~8, right) clearly 
show a significant difference in the electron flux. 
The suitability of a smooth cutoff function would require a spectral fit 
in the 10-30 keV range for this particular example. This example
illustrates that the electron flux or the nonthermal energy calculated
with a smooth cutoff function would yield a significantly different
value than the sharp cutoff function.
Smooth functions appear to be more realistic in a collisional plasma than an infinitely sharp edge at the low-energy cutoff.

\subsection{	Uncertainties of Nonthermal Energies in Flares	}

A central question of this study is the statistical uncertainty of the
various forms of flare energies, in particular the nonthermal
energies of flares. From the distributions of (logarithmic)
nonthermal energies we found means and standard deviations
of $q_E=0.40 \pm 0.10$ for the electron number model (Fig.~6a),
$q_E=0.58 \pm 0.16$ for the time-of-flight model (Fig.~6c),
and $q_E=0.45 \pm 0.19$ for the warm target model (Fig.~6b),
which are fully compatible with the previously measured values of
$q_E=0.51 \pm 0.17$ based on the warm target model by using
different temperature mixtures (Aschwanden et al.~2017).
The error of the mean is even smaller when all measurements
from the three methods are combined, i.e., $q_e=0.57\pm0.08$
(Fig.~7d). However, the standard deviations of the energy
ratios scatter by factors of $\sigma \approx 8-24$ (Fig.~6),
which represents the uncertainties for single events.
Combining the first three methods, the uncertainty for
a single event comes down to a factor of $\sigma=5.4$ (Fig.~7d).
Since these energy ratios $q_E=E_{nth}/E_{diss}$ involve both
the nonthermal energies and the dissipated magnetic energies,
the uncertainties of both types of energies are folded into
these uncertainties. In summary, we can say that the
statistical error of the mean nonthermal-to-magnetic energy ratio
has been reduced to $\gapprox 8\%$, while the uncertainty
of the ratio for an individual event has been reduced to
a factor of 5. Future studies should concentrate on cases with
unphysical values, such as flares that yield nonthermal
energies that are larger than the dissipated magnetic energy.

\section{	CONCLUSIONS					}

In this study we revisit the nonthermal flare energies, previously
calculated for 191 flare events (of M and X-class)
observed with RHESSI during the time period of 2010-2014
(Aschwanden et al.~2016), based on the warm target model of
Kontar et al.~(2015, 2019). The warm target model predicts
a low-energy cutoff that scales linearly with the temperature
$T_e$ of the warm target and the spectral power-law slope
$\delta$ of the nonthermal electron flux, i.e.,
$\varepsilon_c \approx \delta k_B T_e$ (Kontar et al.~2015).
The power-law slope $\delta$ is obtained from a spectral fit
of RHESSI data with the OSPEX software, applied to the
nonthermal energy range of $\varepsilon \approx 10-30$ keV.
The temperature is weighted by a mixture of preflare plasma
temperatures ($T_{cold}$) and heated upflowing evaporating
flare plasma temperatures ($T_{hot}$), which has a mean value
of $T_e=8.8 \pm 6.0$ MK for AIA data,
from which the mean values of the differential emission measure
(DEM) peak temperatures were used in the previous study
(Aschwanden et al.~2016). These parameters yielded a mean
energy cutoff of $e_{wt}=6.2 \pm 1.6$ keV in the warm target
model, and an energy fraction of $q_E=0.51 \pm 0.17$ for
the mean (logarithmic) ratio of the nonthermal energy to
the dissipated magnetic energy.

Since the nonthermal flare energies represent the largest
fraction of the total energy budget in flares, and since
the determination of the nonthermal flare energy has the
largest uncertainty due to the unknown low-energy cutoff,
we decided to revisit the calculation of nonthermal energies
with four different physical models that complement each
other, which we summarize in the following.

\begin{enumerate}
\item{The {\sl electron number model} estimates the number
of electrons (in the preflare phase) that can be accelerated
in a flaring region, which is the product of the (preflare)
electron density $n_e$, the flare volume $V$, 
and the flare duration $\tau_{flare}$.
Some geometry factor is required to relate the acceleration
volume to the flaring volume seen in EUV. Setting this total
electron number equal to the total number of electrons contained
in the electron injection spectrum according to the thick-target
model, a low-energy cutoff $\varepsilon_{en}$ can be derived
that depends on the spectral parameters $[I_1(t), \gamma(t)]$
and the physical parameters $[n_e, V, \tau_{flare}]$.
Using this model we infer a low-energy
cutoff of $\varepsilon_{en}=10.8 \pm 7.5$ keV 
and a value of $E_{nth}/E_{diss}=0.40 \pm 0.10$ for the ratio of the
nonthermal to the dissipated magnetic energy.}

\item{The {\sl time-of-flight} model assumes the equivalence
of the collisional deflection time $t_{defl}$ and the electron
time-of-flight time scale $t_{tof}$. This model essentially
assumes a non-collisional plasma for $t_{tof} < t_{defl}$,
and a collisional plasma for longer propagation times,
$t_{tof} > t_{defl}$. This model predicts a low-energy
cutoff that depends on the electron time-of-flight distance
$L_{tof}$ (which we approximate with the length scale
$L_{tof}$ of the flare area) and the preflare electron
density $n_e$. Using this model we infer a low-energy
cutoff of $\varepsilon_{tof}=9.8 \pm 9.5$ keV and a value of
$E_{nth}/E_{diss}=0.58 \pm 0.16$ for the ratio of the
nonthermal to the dissipated magnetic energy.}

\item{The {\sl warm target} model, derived by
Kontar et al.~(2015, 2016), replaces the original cold
thick-target model, where in addition to the
``cold'' chromospheric plasma, a ``warm'' coronal 
plasma is added, where the precipitating electrons
collisionally thermalize in the ambient
coronal Maxwellian distribution. 
Importantly, the thermalized electrons
contribute to the overall thermal spectrum.
The ``warm'' temperature of the coronal plasma can be a mixture
of cold and hot plasma, which we approximate here with the geometric
mean of the ``cold'' EUV temperature (obtained from DEM modeling)
and the ``hot'' soft X-ray plasma temperature (obtained from RHESSI
fitting with a combined thermal plus nonthermal spectrum).
Using this model we infer a low-energy
cutoff of $\varepsilon_{wt}=9.9 \pm 4.8$ keV and a value of
$E_{nth}/E_{diss}=0.45 \pm 0.10$ for the ratio of the
nonthermal to the dissipated magnetic energy.}

\item{The {\sl spectral cross-over} model is included here
for comparison. An upper limit for the low-energy cutoff
can be found from the intersection point of the thermal
(low-energy) component and the nonthermal (high-energy) 
component in spectral fits of RHESSI data, using the OSPEX software.
As it was established earlier, the low-energy cutoff
is different by about a factor of two, for which
we find a range of $\varepsilon_{co}=21.3 \pm 5.8$ keV.}
\end{enumerate}

In summary, we conclude that the first three models yield
consistent values for the low-energy cutoff in the
order of $\varepsilon \approx 10$ keV, while the spectral
cross-over model yields upper limits only, at $\varepsilon
\approx 21$ keV. It is interesting that the first three different models
with different assumptions lead to similar results.
Combining all three methods,  we find a mean energy partition
of $q_E=0.57 \pm 0.08$ for nonthermal energies,
while the uncertainty in a single event has been reduced
to a factor of 5.

\section*{	APPENDIX A: Smooth Low-Energy Cutoff Function		}

The electron injection spectrum in the thick-target model
is generally rendered with a power-law function that drops off
steeply above the cutoff energy at $\varepsilon \ge \varepsilon_c$,
and is set to zero below this cutoff energy at $\varepsilon
< \varepsilon_c$ (e.g., Holman 2003). This form of a spectral
function results into an extremely
narrow function in energy that is almost mono-energetic.
For collisional processes, a sharp cutoff function may be
unrealistic, while a smooth cutoff function is more likely
to occur. We define a smooth electron injection function
$f_e(\varepsilon)$ simply by introducing a multiplicative term
$(1 - \varepsilon_c/\varepsilon)$,
$$
	f_e^{sm}(\varepsilon) =
	f_e(\varepsilon) \left( 1 - {\varepsilon_{min} \over \varepsilon }
	\right) \ ,
	\eqno(A1)
$$
which fulfills the two constraints of a low-energy cutoff of
$f_e^{sm}(\varepsilon=\varepsilon_{min}) = 0$ 
and the approximative
form of a power-law-like function at higher energies, i.e.,
at $\varepsilon \gapprox \varepsilon_{\min}$.

The smooth electron injection spectrum (as shown with thick linestyle
in Fig.~8) has then the functional form of (based on Eq.~2),
$$
	f_e^{sm}(\varepsilon) =
	f_1 \left( {\varepsilon \over \varepsilon_{min} } \right)^{-\delta}
	\left( 1 - {\varepsilon_{min} \over \varepsilon }
	\right) \ .
	\eqno(A2)
$$
The smoothed electron injection function has a minimum energy of
$\varepsilon_{min}$, and a peak at $\varepsilon_{peak}$. If we
set the peak energy equal to the sharp cutoff, i.e.,
$\varepsilon_{peak}=\varepsilon_c$, which can be calculated from
the derivative $\partial f_e^{sm}(\varepsilon ) / \partial \varepsilon=0$,
we obtain
$$
	\varepsilon_{peak}=\varepsilon_{min} \left( 1 + {1 \over \delta}
	\right) \ .
	\eqno(A3)
$$
For instance, for the example shown in Fig.~8, the energy ratio is
$\varepsilon_{peak}=\varepsilon_{min} ( 1 + 1/\delta ) = 4/3 = 1.333$
for $\delta = 3$. For steeper slopes $\delta$ the ratio becomes smaller,
such as $\varepsilon_{peak}=\varepsilon_{min} ( 1 + 1/\delta) = 9/8 = 1.125$
for $\delta=8$.

We can now analytically calculate
the functional form of the total number of electrons above
a cutoff energy of $\varepsilon_c$,
$$
	F^{sm}(\varepsilon \ge \varepsilon_{min})
	= \int_{{\varepsilon}_{min}}^\infty f_e(\varepsilon)\
	\left( 1 - {\varepsilon_c \over \varepsilon } \right)
	d\varepsilon
	= F(\varepsilon \ge \varepsilon_{c})
	\left({1 \over 1 + \gamma}\right)
	\qquad [{\rm electrons}\ {\rm s}^{-1}] \ ,
	\eqno(A4)
$$
where the integration of $F(\varepsilon \ge \varepsilon_c)$
produces a simple multiplication factor $1 / (1 + \gamma)$ that
depends on the spectral slope $\gamma$ of the photon spectrum only.

Similarly, we can analytically calculate the power
$F^{sm}(\varepsilon \ge \varepsilon_c)$ in nonthermal
electrons above this cutoff energy,
$$
	P^{sm}(\varepsilon \ge \varepsilon_{min})
	= \int_{{\varepsilon}_c}^\infty f_e(\varepsilon)\
	\varepsilon \
	\left( 1 - {\varepsilon_c \over \varepsilon } \right)
	d\varepsilon
	= P(\varepsilon \ge \varepsilon_c)\ {1 \over \gamma}
	\qquad [{\rm erg }\ {\rm s}^{-1}] \ ,
	\eqno(A5)
$$
where the integration of $P(\varepsilon \ge \varepsilon_c)$
produces a similar multiplication factor $(1 / \gamma)$ that
depends on the spectral slope $\gamma$ of the photon spectrum only.
Since the correction of the sharply peaked electron injection
function by a smoothed function depends on the power law slope
$\gamma$, we expect a change in the energy-dependence of the
distribution functions.

The smooth definition of the electron injection function affects
also the value of the low-energy cutoff
for the electron number model, since the total number
of electrons $N_e$ (Eq.~6) changes as,
$$
	N_e = F(\varepsilon \ge \varepsilon_c)
	\left({ 1 \over 1 + \gamma }\right) \ \tau_{flare}
	\qquad [{\rm electrons}] \ ,
	\eqno(A6)
$$
and the resulting low-energy cutoff is modified by the factor
$1/(1+\gamma)$, compared with Eq.~(9), i.e.,
$$
        \varepsilon_{en} = \epsilon_1
        \left[ n_{e0}\ L_{10}^3\ q_{geo}\ \epsilon_1 \over
        0.72 \ \gamma^2 (\gamma^2-1) \ I_1\ \tau_{flare}
	\right]^{-1/\gamma}
        \qquad [{\rm keV}]\ .
	\eqno(A7)
$$
Thus, the smooth electron injection function causes this modification
in the calculation of the low-energy cutoff of the electron number model,
but it does not affect the time-of-flight model (Eq.~20), the warm target
model (Eq.~28), or the cross-over model (Eq.~33), since these other
models do not directly depend on the chosen electron injection function.
Future studies may fit the smoothed electron injection function (Eq.~A1)
in order to obtain a more accurate estimate of flare energies.

\section*{	APPENDIX B: Parameter Dependence of Low-Energy Cutoff 	}

\subsection*{B1 : The Electron Number Model}

The input parameters of our low-energy cutoff models affect the
final result of the low-energy cutoff value $\varepsilon$ in a specific way 
for each parameter. In Table 1 (based on the parameter distributions shown
in Fig.~3) we list the mean and standard deviations $x_{mean}\pm\sigma$
of each observed variable ($x=T_e, \gamma, L, \tau_{flare}, EM, I_1, n_e, n_{e0}$),
which can be characterized by the variance ratio $\sigma/x_{mean}$,
found to range from $\sigma_{\gamma}/\gamma=1.20$ (for spectral
slopes) up to a factor of $\sigma_{EM}/EM=11.0$ (for emission measures)
(Table 1). 

We investigate now how these typical parameter variations affect the
typical values of the resulting low-energy cutoffs $\varepsilon$.
We start with the electron number model (Eq.~9),
$$
        \varepsilon_{en} = \epsilon_1
        \left[ n_{e0}\ L_{10}^3\ q_{geo}\ \epsilon_1 \over
        0.72 \ \gamma^2 (\gamma-1)\ I_1\ \tau_{flare} \right]^{-1/\gamma}
        \qquad [{\rm keV}]\ .
	\eqno(B1)
$$
Denoting the mean value of the preflare electron density with $n_{e0}$
and the value of a standard deviation higher with $\tilde{n}_{e0}$
(with $[{\tilde{n}_{e0} / n_{e0} }]=6.34$ according to Table 1),
the corresponding low-energy cutoff value $\tilde{\varepsilon}_{en}$ is
$$
	{\tilde{\varepsilon}_{en} \over \varepsilon_{en}} =  
	\left[ {\tilde{n}_{e0} \over n_{e0} } \right]^{-1/\gamma} =
	\left[ 6.34 \right]^{-1/7} = 0.77 \ ,
	\eqno(B2)
$$
which means that the low-energy cutoff value $\tilde{\varepsilon}_{en}$
comes out to be 23\% lower for a preflare electron density that is
a standard deviation higher than the mean value. This value can
be considered as an upper limit of the uncertainty of the low-energy
cutoff value, if all the variance in the electron density measurements
are due to measurement errors in the electron density. Practically, since the
obtained mean value is $\varepsilon_{en}=10.8\pm7.5$ keV (Fig.~4a), this
uncertainty is $0.23 \times 10.8$ keV $\approx 2.5$ keV.

Next we investigate the uncertainty caused by the nonthermal flux $I_1$.
Denoting the mean value of the nonthermal flux $I_1$
and the value of a standard deviation higher with $\tilde{I}_1$
(with $[\tilde{I}_1 / I_1] = 6.40$ according to Table 1),
the corresponding low-energy cutoff value $\tilde{\varepsilon}_{en}$ is
$$
	{\tilde{\varepsilon}_{en} \over \varepsilon_{en}} =  
	\left[ {I_1 \over \tilde{I}_1 } \right]^{-1/\gamma} =
	\left[ 1. / 6.40 \right]^{-1/7} = 1.30 \ ,
	\eqno(B3)
$$
which means that the low-energy cutoff value $\tilde{\varepsilon}_{en}$
comes out to be 30\% higher for a nonthermal flux that is
a standard deviation higher than the mean value. This value indicates
an uncertainty of $0.30 \times 10.8$ keV $\approx$ 3.2 keV, which is
an upper limit of the uncertainty, if all variance is due to
measurement errors of the nonthermal flux. 

Next we investigate the uncertainty due to the flare duration 
$\tau_{flare}$.  Denoting the mean value of the flare duration $\tau_{flare}$
and the value of a standard deviation higher with $\tilde{\tau}_{flare}$
(with $[\tilde{\tau}_{flare} / \tau_{flare}] = 1.84$ according to Table 1),
the corresponding low-energy cutoff value $\tilde{\varepsilon}_{en}$ is
$$
	{\tilde{\varepsilon}_{en} \over \varepsilon_{en}} =  
	\left[ {\tau_{flare} \over \tilde{\tau}_{flare}} \right]^{-1/\gamma} =
	\left[ {1 \over 1.84} \right]^{-1/7} = 1.15 \ ,
	\eqno(B4)
$$
which means that the low-energy cutoff value $\tilde{\varepsilon}_{en}$
comes out to be 15\% higher for a flare duration that is
a standard deviation higher than the mean value. This value indicates
an uncertainty of $0.15 \times 10.8$ keV $\approx$ 1.6 keV, which is
an upper limit of the uncertainty, if all variance is due to
measurement errors of the flare duration. 

Next we investigate the uncertainty due to the flare length scale 
$L$.  Denoting the mean value of the length scale $L$
and the value of a standard deviation higher with $\tilde{L}$
(with $[\tilde{L} / L] = 1.55$ according to Table 1),
the corresponding low-energy cutoff value $\tilde{\varepsilon}_{en}$ is
$$
	{\tilde{\varepsilon}_{en} \over \varepsilon_{en}} =  
	\left[ {\tilde{L}^3 \over L^3} \right]^{-1/\gamma} =
	\left[ 1.55^3 \right]^{-1/7} = 0.83 \ ,
	\eqno(B5)
$$
which means that the low-energy cutoff value $\tilde{\varepsilon}_{en}$
comes out to be 17\% lower for a length scale that is
a standard deviation larger than the mean value. This value indicates
an uncertainty of $0.17 \times 10.8$ keV $\approx$ 1.8 keV, which is
an upper limit of the uncertainty, if all variance is due to
measurement errors of the flare length scale. 

Next we investigate the uncertainty due to the spectral slope 
$\gamma$.  Denoting the mean value of the spectral slope $\gamma$
and the value of a standard deviation higher with $\tilde{\gamma}$
(with $[\tilde{\gamma} / \gamma] = 1.20$ according to Table 1),
the corresponding low-energy cutoff value $\tilde{\varepsilon}_{en}$ is
$$
	{\tilde{\varepsilon}_{en} \over \varepsilon_{en}} =  
	\left[ { \gamma^2 (\gamma-1) \over \tilde{\gamma}^2 
	(\tilde{\gamma}-1)} \right]^{-1/\gamma} =
	\left[ 1.63 \right]^{-1/7} = 0.93 \ ,
	\eqno(B6)
$$
which means that the low-energy cutoff value $\tilde{\varepsilon}_{en}$
comes out to be 7\% lower for a spectral index scale that is
a standard deviation larger than the mean value. This value indicates
an uncertainty of $0.07 \times 10.8$ keV $\approx$ 0.8 keV, which is
an upper limit of the uncertainty, if all variance is due to
measurement errors of the spectral slope. 

Finally, we investigate also the uncertainty due to the geometric parameter
$q_{geo}=1/4$, which is assumed for the ratio of the flare arcade volume
with respect to an encompassing cube.
Denoting the mean value of the geometry factor with $q_{geo}$
and the value of a factor two higher with $\tilde{q}_{geo}$
(i.e., $[\tilde{\gamma} / \gamma] = 2$),
the corresponding low-energy cutoff value $\tilde{\varepsilon}_{en}$ is
$$
	{\tilde{\varepsilon}_{en} \over \varepsilon_{en}} =  
	\left[ { \tilde{q}_{geo} \over q_{geo}} \right]^{-1/\gamma} =
	\left[ 2 \right]^{-1/7} = 0.91 \ ,
	\eqno(B7)
$$
which means that the low-energy cutoff value $\tilde{\varepsilon}_{en}$
comes out to be 9\% lower for a geometry factor that is
a factor two larger than the mean value. This value indicates
an uncertainty of $0.09 \times 10.8$ keV $\approx$ 1.0 keV, which is
an upper limit on the uncertainty of the geometry factor.

In summary, upper limits of the uncertainties $\sigma_x$ of the 
low-energy cutoff $\varepsilon_{en}$ in our electron number model 
are estimated (in decreasing order) from the following parameters: 
the nonthermal flux $I_1$ (i.e., $\sigma_{I1}< 30\%$ of the low-energy cutoff value),)
preflare electron density $n_{e0} (< 23\%)$,
flare length scale $\tau_{flare} (< 17\%)$,
flare duration $\tau_{flare} (< 15\%)$,
geometric model $q_{geo} (< 9\%)$, and
spectral index $\gamma (< 7\%)$. 
In these estimates we make the assumption
that the variance of the values is entirely caused by measurement errors,
which constitutes upper limits on the uncertainties of the low-energy cutoff values.

\subsection*{B2 : The Time-of-Flight Model}

We proceed now to our second model, the so-called time-of-flight model,
which depends on two parameters only, the length scale $L$ and the mean
electron density $n_e$ during flares (Eq.~20),
$$
	e_{tof} \approx 28
	\left( {L \over 10^{10}\ {\rm cm}} \right)^{1/2}	
	\left( {n_e \over 10^{10}\ {\rm cm}^{-3}} \right)^{1/2}	
	\qquad {\rm [keV]}\ .
	\eqno(B8)
$$
Similarly to the previous method, we investigate the uncertainty due to the 
length scale $L$. Denoting the mean value of the length scale $L$
and the value of a standard deviation higher with $\tilde{L}$
(with $[\tilde{L} / L] = 1.55$ according to Table 1),
the corresponding low-energy cutoff value $\tilde{\varepsilon}_{en}$ is
$$
	{\tilde{\varepsilon}_{en} \over \varepsilon_{en}} =  
	\left[ {\tilde{L} \over L} \right]^{1/2} =
	\left[ 1.55 \right]^{1/2} = 1.24 \ ,
	\eqno(B9)
$$
which means that the low-energy cutoff value $\tilde{\varepsilon}_{en}$
comes out to be 24\% higher for a length scale that is
a standard deviation larger than the mean value. 
Using the distribution shown in Fig.~4b, i.e., $\varepsilon_{tof}
=9.8 \pm 9.5$ keV. This value indicates
a mean uncertainty of $0.24 \times 9.8$ keV $\approx$ 2.4 keV, which is
an upper limit of the uncertainty, when all variance is due to
measurement errors of the length scale. 

Denoting the mean value of the flare electron density with $n_{e}$
and the value of a standard deviation higher with $\tilde{n}_{e}$
(with $[{\tilde{n}_{e} / n_{e} }]=3.69$ according to Table 1),
the corresponding low-energy cutoff value $\tilde{\varepsilon}_{en}$ is
$$
        {\tilde{\varepsilon}_{en} \over \varepsilon_{en}} =
        \left[ {\tilde{n}_{e} \over n_{e} } \right]^{1/2} =
        \left[ 3.69 \right]^{1/2} = 1.92 \ ,
	\eqno(B10)
$$
which means that the low-energy cutoff value $\tilde{\varepsilon}_{en}$
comes out to be 92\% higher for a flare electron density that is
a standard deviation higher than the mean value. This value can
be considered as an upper limit of the uncertainty of the low-energy
cutoff value, if all the variance in the electron density measurements
are due to measurement errors in the electron density. Practically, since the
obtained mean value is $\varepsilon_{en}=9.8\pm9.5$ keV (Fig.~4b), this
uncertainty is $0.92 \times 10.8$ keV $\approx 9.0$ keV. This large
uncertainty implies a high sensitivity of the low-energy cutoff on the
flare density, while it substantially less sensitive to the flare
length scale. It is therefore imperative to measure the flare density
accurately, which requires detailed DEM analysis.

\subsection*{B3 : The Warm Target Model}

Finally, we investigate the parameter dependence of the warm target model,
which in its simplest form (Eq.~28),
$$
	\varepsilon_{wt} \approx (\xi + 2) k_B T_e = \delta \ k_B T_e \ 
	= (\gamma + 1) k_B T_e,
	\eqno(B12)
$$
where $\xi = \gamma - 1$ is the power-law slope of the
source-integrated mean electron flux spectrum (see Eqs.~(8)-(10)
in Kontar et al.~2015), and $T_e$ is the temperature of the
warm target plasma. 
Denoting the mean value of the spectral index with $\gamma$
and the value of a standard deviation higher with $\tilde{\gamma}$
(with $[\tilde{\gamma} / \gamma]=1.20$ according to Table 1),
the corresponding low-energy cutoff value $\tilde{\varepsilon}_{en}$ is
$$
        {\tilde{\varepsilon}_{en} \over \varepsilon_{en}} =
        \left[ {\tilde{\gamma} + 1 \over \gamma + 1 } \right] = 1.18
	\eqno(B13)
$$
which means that the low-energy cutoff value $\tilde{\varepsilon}_{en}$
comes out to be 18\% higher for a spectral index that is
a standard deviation higher than the mean value. This value can
be considered as an upper limit of the uncertainty of the low-energy
cutoff value, if all the variance in the spectral index measurements
are due to measurement errors in the electron density. Practically, since the
obtained mean value is $\varepsilon_{en}=9.9\pm4.8$ keV (Fig.~4c), this
uncertainty is $0.18 \times 9.9$ keV $\approx 2.0$ keV. 

The temperature dependence can be calculated by  
denoting the mean value of the spectral index with $T_e$
and the value of a standard deviation higher with $\tilde{T}_e$
(with $[\tilde{T}_e / T_e]=1.40$ according to Table 1),
the corresponding low-energy cutoff value $\tilde{\varepsilon}_{en}$ is
$$
        {\tilde{\varepsilon}_{en} \over \varepsilon_{en}} =
        \left[ {\tilde{T}_e \over T_e} \right] = 1.40
	\eqno(B14)
$$
which means that the low-energy cutoff value $\tilde{\varepsilon}_{en}$
comes out to be 40\% higher for a spectral index that is
a standard deviation higher than the mean value. This value can
be considered as an upper limit of the uncertainty of the low-energy
cutoff value, if all the variance in the spectral index measurements
are due to measurement errors in the electron density. Practically, since the
obtained mean value is $\varepsilon_{en}=9.9\pm4.8$ keV (Fig.~4c), this
uncertainty is $0.40 \times 9.9$ keV $\approx 4.0$ keV. 

Thus, for the warm target model, uncertainties up to 18\% of the low-energy
cutoff could be arise due to uncertainties in the spectral index, and 
uncertainties up to 40\% of the low-energy cutoff could be caused by 
uncertainties of the temperature measurement. 

\bigskip
\acknowledgements
We acknowledge useful discussions with John Raymond.
This work was partially supported by NASA contracts NNX11A099G,
NNG04EA00C (SDO/AIA), and NNG09FA40C (IRIS).
EPK and NLSJ were supported by the Science and Technology Facilities 
Council (STFC) Consolidated Grant ST/L000533/1.

\clearpage
\section*{References} % REFERENCES
\def\ref#1{\par\noindent\hangindent1cm {#1}}

\ref{Aschwanden, M.J.: 2004, {\sl Physics of the Solar Corona.
	An Introduction}, Berlin: Springer and Praxis, p.216.}
\ref{Aschwanden, M.J. 2007, ApJ 661, 1242.}
\ref{Aschwanden, M.J., Xu, Y., and Jing, J. 2014, ApJ 797, 50.
        {\sl Global energetics of solar flares: I. Magnetic Energies}}
\ref{Aschwanden, M.J. 2015, ApJ 814, 19.}
\ref{Aschwanden, M.J., Boerner, P., Ryan, D., Caspi, A., McTiernan, J.M.,
        and Warren, H.P., 2015, ApJ 802, 53.
        {\sl Global energetics of solar flares: II. Thermal Energies}}
\ref{Aschwanden, M.J., O'Flannagain, A., Caspi, A., McTiernan, J.M.,
        Holman, G., Schwartz, R.A., and Kontar, E.P. 2016, ApJ 832, 27.
        {\sl Global energetics of solar flares: III. Nonthermal Energies}}
\ref{Aschwanden, M.J. 2016, ApJ 831, 105.
        {\sl Global energetics of solar flares. IV. Coronal
        mass ejection energetics}}
\ref{Aschwanden, M.J., Caspi, A., Cohen, C.M.S., Holman, G.D., Jing, J.,
        Kretzschmar, M., Kontar, E.P., McTiernan, J.M., O'Flannagain, A.,
        Richardson, I.G., Ryan, D., Warren, H.P., and Xu,Y. 2017, ApJ 836, 17.
        {\sl Global energetics of solar flares: V. Energy closure}}
\ref{Aschwanden, M.J. 2017, ApJ 847, 27.
        {\sl Global energetics of solar flares. VI. Refined
        energetics of coronal mass ejections.}}
\ref{Aschwanden, M.J. and Gopalswamy, N. 2019, ApJ (in press).
        {\sl Global energetics of solar flares. VII. The aerodynamic
	drag in coronal mass ejections.}}
\ref{Aschwanden, M.J. 2019, {\sl New Millennium Solar Physics},
        Astrophysics and Space Science Library Vol.~458,
        ISBN 978-3-030-13954-4; New York: Springer.}
\ref{Aschwanden, M.J. 2019, {\sl New Millennium Solar Physics},
	Section 11.8, New York: Springer, (in press),
	http://www.lmsal.com/$\sim$aschwand/bookmarks$\_$books2.html.}
\ref{Benz, A.O. 1993, {\sl Plasma Astrophysics. Kinetic processes in
	solar and stellar coronae}, Kluwer Academic Publishers: Dordrecht, p.45.}
\ref{Bian, N.H., Emslie, A.G., Stackhouse, D.J., and Kontar, E.P. 2014,
	ApJ 796, 142.}
\ref{Brown, J.C. 1971, SoPh 18, 489.}
\ref{Brown, J.C. 1974, in Proc. Symp. IAU COll. 57, {\sl Coronal
	disturbances}, (e.d., G.J. Newkirk, Jr. (Dordrecht: Reidel), 523.}
\ref{Carmichael, H. 1964, in {\sl The Physics of Solar Flares},
	Proc. AAS-NASA Symposium, (ed. W.N. Hess), NASA-SP 50,
	NASA Science and Technical Information Division, Washington DC p.451.}
\ref{Crosby, N.M., Aschwanden, M.J., and Dennis, B. 1993, SoPh 143, 275.}
\ref{Culhane, J.L. 1969, MNRAS 144, 375.}
\ref{Culhane, J.L. and Acton, L. 1970, MNRAS 151, 141.}
\ref{Dennis, B.R. 1985, SoPh 100, 465.}
\ref{Dulk, G.A. and Dennis, B.R. 1982, ApJ 260, 875.}
\ref{Galloway, R.K., MacKinnon, A.L., Kontar, E.P. and Helander P.
	2005, A\&A 438, 1107.}
\ref{Goncharov, P.R., KJuteev, B.V., Ozaki, T., and Sudo, S.
	2010, PhPl 17, 112313.}
\ref{Guo, J., Emslie, A.G., Kontar, E.P., et al.~2012a,
	A\&A 543, A53.}
\ref{Guo, J., Emslie, A.G., Massone, A.M., Piana, M., and PIana, M.
	~2012b, ApJ 755, 32.}
\ref{Guo, J., Emslie, A.G., and Piana, M. 2013, ApJ 766, 28.}
\ref{Hirayama, T. 1974, SoPh 34, 323.}
\ref{Holman, G.D. 2003, ApJ 586, 606.}
\ref{Holman, G.D., Aschwanden, M.J., Aurass, M.J. et al.~2011,
	SSRv 159, 107.}
\ref{Ireland, J., Tolbert, A.K., Schwartz, R.A., Holman, G.D., and
	Dennis, B.R. 2013, ApJ 769, 89.}
\ref{Jeffrey, N.L.S., Kontar, E.P., Bian, N.H., and Emslie, A.G.
	2014, ApJ 787, 86.}
\ref{Jeffrey, N.L.S., Kontar, E.P., and Emslie, A.G. 2015, A\&A 584, 89.}
\ref{Kontar, E.P., Jeffrey, N.L.S., Emslie, A.G., and Bian, N.H.
	2015, ApJ 809, 35.}
\ref{Kontar, E.P., Jeffrey, N.L.S., and Emslie, A.G. 2019,
	ApJ 871, 225.}
\ref{Kopp, G. and Pneuman, G.W. 1976, SoPh 50, 85.}
\ref{Lemen,J.R., Title, A.M., Akin,D. J., Boerner, P.F., 
	Chou, C., Drake, J.F., Duncan, D.W., Edwards, C.G., et al.
 	2012, SoPh 275, 17.}
\ref{Lin, R.P., Dennis,B.R., Hurford,G.J., Smith,D.M., 
	Zehnder,A., Harvey,P.R., Curtis,D.W., et al. 2002, Solar Phys. 210, 3.}
\ref{Pesnell, W.D., Thompson, B.J., and Chamberlin, P.C. 2011, 
 	SoPh 275, 3.}
\ref{Petrosian, V. 2000, SSRv 173, 535.}
\ref{Scherrer, P.H., Schou, J., Bush, R. I., Kosovichev, A.G., 
	Bogart, R.S., Hoeksema, J T., Liu, Y., Duvall, T.L., et al.
 	2012, SoPh 275, 207.}
\ref{Somov, B.V. 2000, {\sl Cosmic Plasma Physics}, Dordrecht: Kluwer
	Academic Publishers.}
\ref{Sturrock, P.A. 1966, Nature 5050, 695.}

%@@@REFERENCES

\clearpage
%%%%%%%%%%%%%%%%%%%%%%%%%%%%%% TABLES  %%%%%%%%%%%%%%%%%%%%%%%%%%%%%%%%%%

\begin{table}
\tabletypesize{\normalsize}
\caption{The ranges $(x_{min}, x_{max})$, medians $(x_{med})$, means and standard deviations
$x_{mean}\pm\sigma$, and variance ratios $(\sigma/x_{mean})$ of the observed variables in the
determination of the low-energy cutoff $\varepsilon_c$ are listed
according to Fig.~2, for a total of 191 M and X-class flares.}
\medskip
\begin{tabular}{lrrrrr}
\hline
Parameter	&Minimum	&Median       &Maximum   &Mean 	        &Variance \\
		&		&	      &          &std           &ratio    \\
		&$x_{min}$	&$x_{med}$    &$x_{max}$ & $x_{mean}\pm\sigma$  & $\sigma/x_{mean}$\\
\hline
Temperature $T_e$ (MK) & 3.4	& 12.5	      & 33.7     & 13.5$\pm$5.4 & 1.40 \\	
Spectral slope $\gamma$& 2.8	& 7.2	      & 10.4     &  7.0$\pm$1.4 & 1.20 \\	
Length scale $L$ (Mm)  & 1.7	& 9.8	      & 34.8     & 10.9$\pm$6.0 & 1.55 \\	
Flare duration $t_{flare}$ (s)
		& $10^{2.20}$     & $10^{3.22}$ & $10^{3.98}$ & $10^{3.2\pm0.27}$ & 1.84 \\
Emission measure $EM$ (cm$^{-3}$)
		& $10^{44.3}$     & $10^{47.0}$ & $10^{51.4}$ & $10^{47.1\pm1.04}$ & 11.0 \\
Photon flux $I_1$ (photons cm$^{-2}$ s$^{-1}$ keV$^{-1}$)
		& $10^{-5.12}$    & $10^{-3.34}$ & $10^{-0.65}$& $10^{-3.27\pm0.81}$& 6.40 \\
Flare electron density $n_{e}$
		& $10^{8.30}$    & $10^{9.97}$ & $10^{12.13}$ & $10^{10.1\pm0.80}$& 3.69 \\
Preflare electron density $n_{e0}$
		& $10^{8.00}$    & $10^{9.34}$ & $10^{13.41}$ & $10^{9.60\pm0.80}$& 6.34 \\
\hline
\end{tabular}
\end{table}

\clearpage
%%%%%%%%%%%%%%%%%%%%%%%%%%%% TABLE 2 %%%%%%%%%%%%%%%%%%%%%%%%%%%%%%

\begin{deluxetable}{rrrrrrrrrrrrr}
\tabletypesize{\normalsize}
\tabletypesize{\footnotesize}
\setlength{\tabcolsep}{0.05in}
\tablecaption{Observables of flare hard X-ray emission in 143
M and X-class flare events.}
\tablewidth{0pt}
\tablehead{
\colhead{ID}&
\colhead{Date}&
\colhead{Time}&
\colhead{GOES}&
\colhead{Heliogr.}&
\colhead{Dur.}&
\colhead{Emission}&
\colhead{Temp.}&
\colhead{Photon}&
\colhead{Spectral}&
\colhead{Length}&
\colhead{Density}&
\colhead{Density}\\
\colhead{}&
\colhead{}&
\colhead{}&
\colhead{class}&
\colhead{position}&
\colhead{flare}&
\colhead{measure}&
\colhead{max.}&
\colhead{flux}&
\colhead{slope}&
\colhead{scale}&
\colhead{maximum}&
\colhead{preflare}\\
\colhead{}&
\colhead{}&
\colhead{}&
\colhead{}&
\colhead{}&
\colhead{$\tau_{flare}$}&
\colhead{$EM_{49}$}&
\colhead{$T_e$}&
\colhead{$I_1$}&
\colhead{$\gamma$}&
\colhead{$L$}&
\colhead{$n_{10}$} &
\colhead{$b_{10}$}\\
\colhead{}&
\colhead{}&
\colhead{}&
\colhead{}&
\colhead{}&
\colhead{(s)}&
\colhead{(cm$^{-3}$)}&
\colhead{(MK)}&
\colhead{(cm$^{2}$ s keV$^{-1}$)}&
\colhead{}&
\colhead{(Mm)}&
\colhead{(cm$^{-3}$)} &
\colhead{(cm$^{-3}$)}\\}
\startdata
   1 & 20100612 & 0030 & M2.0 & N23W47 &    904 &    0.00428 &   10.73 &    0.0000736 &    4.24 &   13.23 &     0.43 &     0.18\\
   2 & 20100613 & 0530 & M1.0 & S24W82 &   1852 &    0.00002 &   12.67 &    0.0000181 &    5.04 &   12.25 &     0.03 &     0.01\\
   3 & 20100807 & 1755 & M1.0 & N13E34 &   3700 &    0.00005 &   12.03 &    0.0007735 &    4.09 &   25.10 &     0.02 &     1.64\\
   4 & 20101016 & 1907 & M2.9 & S18W26 &   1572 &    0.03886 &   18.18 &    0.0004585 &    8.06 &   15.13 &     1.06 &     0.08\\
  10 & 20110213 & 1728 & M6.6 & S21E04 &   2324 &    0.01753 &   19.21 &    0.0088956 &    7.10 &   15.94 &     0.66 &     0.03\\
  12 & 20110215 & 0144 & X2.2 & S21W12 &   2628 &    0.24398 &   21.25 &    0.0443819 &    7.06 &   28.41 &     1.03 &     0.13\\
  13 & 20110216 & 0132 & M1.0 & S22W27 &   1368 &    0.00240 &   18.71 &    0.0007753 &    6.83 &   12.16 &     0.37 &     0.14\\
  15 & 20110216 & 1419 & M1.6 & S23W33 &   1692 &    0.01667 &   12.12 &    0.0003076 &    7.54 &   10.74 &     1.16 &     0.33\\
  16 & 20110218 & 0955 & M6.6 & S21W55 &   1780 &    0.09996 &    9.70 &    0.0083943 &    7.21 &   10.61 &     2.89 &     0.15\\
  18 & 20110218 & 1259 & M1.4 & S20W70 &   1944 &    0.01476 &   13.66 &    0.0006862 &    7.13 &    6.44 &     2.35 &     0.60\\
  19 & 20110218 & 1400 & M1.0 & N17E04 &   1264 &    0.01366 &    6.92 &    0.0002906 &    4.33 &    9.43 &     1.28 &     0.21\\
  20 & 20110218 & 2056 & M1.3 & N15E00 &    884 &    0.01607 &    7.51 &    0.0001562 &    7.99 &    8.43 &     1.64 &     0.13\\
  21 & 20110224 & 0723 & M3.5 & N14E87 &   3332 &    0.01042 &   10.86 &    0.0000111 &    9.23 &   20.02 &     0.36 &     0.06\\
  22 & 20110228 & 1238 & M1.1 & N22E35 &    732 &    0.00133 &    8.45 &    0.0013909 &    6.51 &   10.20 &     0.35 &     0.26\\
  23 & 20110307 & 0500 & M1.2 & N23W47 &   1340 &    0.00166 &    8.11 &    0.0004404 &    7.35 &    5.98 &     0.88 &     0.88\\
  28 & 20110307 & 1943 & M3.7 & N30W48 &   3196 &    0.00172 &   10.61 &    0.0029535 &    5.13 &   26.55 &     0.10 &     0.03\\
  29 & 20110307 & 2145 & M1.5 & S17W82 &   1232 &    0.00071 &   10.31 &    0.0023961 &    5.78 &    5.73 &     0.61 &     0.04\\
  30 & 20110308 & 0224 & M1.3 & S18W80 &   1460 &    0.01306 &    4.13 &    0.0008550 &    6.69 &    9.31 &     1.27 &     0.19\\
  31 & 20110308 & 0337 & M1.5 & S21E72 &   2768 &    8.46763 &    5.67 &    0.0000492 &    8.15 &   23.95 &     7.85 &     0.09\\
  33 & 20110308 & 1808 & M4.4 & S17W88 &    848 &    0.00494 &   22.02 &    0.0023009 &    7.52 &   16.21 &     0.34 &     0.00\\
  34 & 20110308 & 1946 & M1.5 & S19W87 &   6044 &    0.00313 &    8.75 &    0.0000175 &    9.16 &   16.40 &     0.27 &    20.96\\
  37 & 20110309 & 2313 & X1.5 & N10W11 &   1660 &    0.04176 &   13.88 &    0.0776128 &    6.05 &   34.75 &     0.32 &     0.01\\
  38 & 20110310 & 2234 & M1.1 & S25W86 &   1588 &    0.01840 &    7.67 &    0.0001338 &    7.66 &    5.74 &     3.12 &     0.56\\
  40 & 20110314 & 1930 & M4.2 & N16W49 &   2308 &    0.21034 &   10.88 &    0.0041197 &    6.88 &   11.74 &     3.61 &     0.26\\
  41 & 20110315 & 0018 & M1.0 & N11W83 &   1500 &    0.02256 &    8.97 &    0.0011179 &    5.04 &    4.58 &     4.85 &     3.10\\
  46 & 20110422 & 0435 & M1.8 & S19E40 &   3124 &    0.00986 &   12.04 &    0.0006550 &    6.89 &   15.75 &     0.50 &     0.13\\
  48 & 20110528 & 2109 & M1.1 & S21E70 &   2848 &    0.01151 &   11.79 &    0.0002199 &    7.07 &   11.97 &     0.82 &     0.00\\
  50 & 20110607 & 0616 & M2.5 & S22W53 &   3608 &    5.21387 &    7.35 &    0.0019885 &    3.96 &   19.91 &     8.13 &     0.09\\
  51 & 20110614 & 2136 & M1.3 & N14E77 &   2356 &    0.00375 &   10.90 &    0.0002383 &    7.37 &   12.63 &     0.43 &     1.63\\
  52 & 20110727 & 1548 & M1.1 & N20E41 &   2004 &    0.00454 &   11.38 &    0.0000151 &    8.96 &   16.68 &     0.31 &     0.28\\
  53 & 20110730 & 0204 & M9.3 & N16E35 &   1460 &    0.53662 &   17.06 &    0.0063472 &    7.86 &   16.20 &     3.55 &     0.11\\
  55 & 20110803 & 0308 & M1.1 & N15W23 &   2760 &    0.00503 &   12.88 &    0.0002445 &    7.64 &    8.66 &     0.88 &     0.00\\
  61 & 20110809 & 0748 & X6.9 & N20W69 &   2256 &    0.17734 &   25.80 &    0.2225979 &    7.38 &   28.85 &     0.86 &     0.39\\
  63 & 20110905 & 0408 & M1.6 & N18W87 &   1516 &    0.00075 &   14.56 &    0.0000897 &    7.97 &    6.80 &     0.49 &     0.30\\
  64 & 20110905 & 0727 & M1.2 & N18W87 &   2464 &    0.00236 &   14.00 &    0.0000076 &    8.38 &    5.55 &     1.18 &     1.41\\
  65 & 20110906 & 0135 & M5.3 & N15W03 &    692 &    0.02325 &   10.01 &    0.0010473 &    8.42 &   19.15 &     0.58 &     0.05\\
  68 & 20110908 & 1532 & M6.7 & N17W39 &   1764 &    0.11622 &   20.71 &    0.0022988 &    8.36 &   16.92 &     1.55 &     1.16\\
  69 & 20110909 & 0601 & M2.7 & N14W48 &   1644 &    0.02375 &    9.37 &    0.0018086 &    7.23 &   17.19 &     0.68 &     0.09\\
  70 & 20110909 & 1239 & M1.2 & N15W50 &    408 &    0.00262 &   11.99 &    0.0000095 &    9.44 &    8.41 &     0.66 &     1.41\\
  71 & 20110910 & 0718 & M1.1 & N14W64 &   2488 &    0.00082 &   21.01 &    0.0001596 &    7.87 &    9.60 &     0.30 &     0.00\\
  77 & 20110923 & 2348 & M1.9 & N12E56 &   1020 &    0.00323 &   10.26 &    0.0003025 &    7.46 &   15.63 &     0.29 &     0.06\\
  81 & 20110924 & 1719 & M3.1 & N13E54 &   1324 &    0.01758 &    9.39 &    0.0007469 &    7.58 &    7.20 &     2.17 &     0.12\\
  83 & 20110924 & 1909 & M3.0 & N15E50 &   1068 &    0.01551 &    8.75 &    0.0003280 &    7.79 &   23.56 &     0.34 &     0.69\\
  84 & 20110924 & 2029 & M5.8 & N13E52 &   1180 &    0.08850 &    9.40 &    0.0119517 &    5.98 &   11.05 &     2.56 &     0.38\\
  86 & 20110924 & 2345 & M1.0 & S28W66 &   1596 &    0.00126 &   13.44 &    0.0000355 &    7.78 &    6.99 &     0.61 &     0.28\\
  91 & 20110925 & 1526 & M3.7 & N15E39 &    676 &    0.01059 &    8.55 &    0.0001207 &    8.83 &   13.64 &     0.65 &     3.75\\
  98 & 20111002 & 0037 & M3.9 & N10W13 &   3696 &    0.01836 &   12.14 &    0.0005113 &    8.25 &   19.25 &     0.51 &     0.01\\
 100 & 20111020 & 0310 & M1.6 & N18W88 &   1044 &    0.00580 &   19.18 &    0.0003065 &    7.98 &    7.15 &     1.26 &     0.03\\
 101 & 20111021 & 1253 & M1.3 & N05W79 &    760 &    0.02016 &    7.03 &    0.0000893 &    7.04 &    6.49 &     2.72 &     0.02\\
 103 & 20111031 & 1455 & M1.1 & N20E88 &   3980 &    0.00846 &   19.72 &    0.0007398 &    7.08 &    4.23 &     3.34 &     1.09\\
 111 & 20111105 & 1110 & M1.1 & N22E43 &   2392 &    0.00081 &   17.51 &    0.0001009 &    7.74 &    8.28 &     0.38 &     0.21\\
 116 & 20111115 & 0903 & M1.2 & N21W72 &   2448 &    0.00132 &    8.89 &    0.0000964 &    8.20 &    7.31 &     0.58 &     0.00\\
 120 & 20111226 & 0213 & M1.5 & S18W34 &   2812 &    0.01884 &    7.57 &    0.0000523 &    7.93 &   13.91 &     0.84 &     0.38\\
 122 & 20111229 & 1340 & M1.9 & S25E70 &   2368 &    0.00718 &   15.36 &    0.0000902 &    8.09 &   14.63 &     0.48 &     0.08\\
 123 & 20111229 & 2143 & M2.0 & S25E67 &    632 &    0.00215 &   15.03 &    0.0002077 &    7.91 &   11.86 &     0.36 &     0.07\\
 125 & 20111231 & 1309 & M2.4 & S25E46 &   1892 &    0.00399 &   20.88 &    0.0010953 &    7.05 &    8.39 &     0.82 &     2.09\\
 126 & 20111231 & 1616 & M1.5 & S22E42 &   1272 &    0.00323 &   13.83 &    0.0001025 &    8.23 &   11.86 &     0.44 &     0.50\\
 157 & 20120427 & 0815 & M1.0 & N13W26 &    732 &    0.00757 &   11.47 &    0.0000452 &    8.65 &   15.58 &     0.45 &     0.00\\
 158 & 20120505 & 1319 & M1.4 & N11E78 &    200 &    0.00455 &   13.05 &    0.0009992 &    5.76 &    9.13 &     0.77 &     0.10\\
 159 & 20120505 & 2256 & M1.3 & N11E73 &    624 &    0.02909 &   17.36 &    0.0011669 &    6.71 &    7.86 &     2.45 &     0.87\\
 160 & 20120506 & 0112 & M1.1 & N11E73 &   1684 &    0.02905 &    3.53 &    0.0017250 &    5.94 &    6.80 &     3.04 &     3.19\\
 167 & 20120510 & 0411 & M5.7 & N12E19 &   1128 &    0.01389 &   12.02 &    0.0196674 &    3.42 &   15.73 &     0.60 &    12.43\\
 168 & 20120510 & 2020 & M1.7 & N12E10 &   1612 &    0.00354 &   12.89 &    0.0019588 &    6.47 &   11.93 &     0.46 &     0.10\\
 169 & 20120517 & 0125 & M5.1 & N07W88 &   2708 &    0.07451 &   11.12 &    0.0002291 &    7.96 &   31.30 &     0.49 &     0.54\\
 170 & 20120603 & 1748 & M3.3 & N15E33 &    852 &    0.08183 &    3.70 &    0.0009645 &    4.13 &   17.31 &     1.26 &     2.42\\
 173 & 20120609 & 1645 & M1.8 & S16E76 &   1724 &    0.01346 &    7.86 &    0.0002785 &    8.03 &    7.50 &     1.79 &     0.09\\
 176 & 20120614 & 1252 & M1.9 & S19E06 &   9628 &    0.02703 &   11.25 &    0.0011941 &    4.24 &    6.13 &     3.43 &     2.87\\
 178 & 20120629 & 0913 & M2.2 & N15E37 &    696 &    0.03472 &   10.61 &    0.0001820 &    7.65 &    8.67 &     2.31 &     0.21\\
 182 & 20120702 & 0026 & M1.1 & N15E01 &   1356 &    0.00326 &   12.41 &    0.0001100 &    7.72 &   10.32 &     0.54 &     0.85\\
 187 & 20120704 & 0947 & M5.3 & S17W18 &   2416 &    0.02938 &   13.49 &    0.0078143 &    7.05 &   10.47 &     1.60 &     0.93\\
 189 & 20120704 & 1435 & M1.3 & S18W20 &    428 &    0.02698 &   12.00 &    0.0022213 &    3.38 &    7.08 &     2.76 &     1.17\\
 190 & 20120704 & 1633 & M1.8 & N14W33 &    828 &    0.01311 &   12.15 &    0.0041792 &    2.76 &   19.31 &     0.43 &     3.14\\
 195 & 20120705 & 0325 & M4.7 & S18W29 &   1768 &    0.03276 &    9.85 &    0.0114881 &    6.97 &    8.49 &     2.31 &     0.75\\
 196 & 20120705 & 0649 & M1.1 & S17W29 &   1208 &    0.00287 &   11.82 &    0.0002549 &    7.40 &    8.11 &     0.73 &     0.30\\
 199 & 20120705 & 1139 & M6.1 & S18W32 &   1056 &    0.02275 &   12.28 &    0.0028190 &    6.09 &   15.74 &     0.76 &     0.24\\
 200 & 20120705 & 1305 & M1.2 & S18W36 &   1400 &    0.00002 &   17.10 &    0.0003799 &    4.58 &   13.83 &     0.03 &     5.73\\
 203 & 20120706 & 0137 & M2.9 & S18W43 &   2748 &    0.02383 &   12.65 &    0.0007113 &    8.18 &    8.49 &     1.97 &     0.22\\
 205 & 20120706 & 0817 & M1.5 & S12W48 &   1392 &    0.01546 &   14.20 &    0.0027188 &    5.80 &    6.86 &     2.19 &     3.25\\
 208 & 20120706 & 1848 & M1.3 & S15E88 &   1348 &    0.00546 &   14.39 &    0.0008365 &    7.04 &   10.17 &     0.72 &     0.43\\
 210 & 20120707 & 0310 & M1.2 & S17W55 &   1664 &    0.00597 &   18.70 &    0.0009195 &    6.95 &    8.67 &     0.96 &     0.10\\
 211 & 20120707 & 0818 & M1.0 & S16E76 &    684 &    0.00182 &   15.12 &    0.0000672 &    6.89 &    5.01 &     1.20 &     1.42\\
 212 & 20120707 & 1057 & M2.6 & S17W59 &    520 &    0.01474 &   21.63 &    0.0022574 &    7.19 &    9.37 &     1.34 &    50.14\\
 214 & 20120708 & 0944 & M1.1 & S16W70 &    768 &    0.00198 &   16.29 &    0.0001030 &    8.15 &    8.49 &     0.57 &     0.00\\
 215 & 20120708 & 1206 & M1.4 & S16W72 &    160 &    0.01743 &   14.30 &    0.0029128 &    6.26 &    6.38 &     2.59 &     1.67\\
 219 & 20120710 & 0605 & M2.0 & S16E30 &   1848 &    0.00205 &   18.43 &    0.0006706 &    7.18 &    9.37 &     0.50 &     0.50\\
 223 & 20120719 & 0417 & M7.7 & S20W88 &   8532 &    0.11691 &   11.72 &    0.0023355 &    6.38 &   17.69 &     1.45 &     0.01\\
 228 & 20120806 & 0433 & M1.6 & S14E88 &    728 &    0.04923 &    6.39 &    0.0022234 &    5.19 &    4.33 &     7.79 &     1.42\\
 230 & 20120817 & 1312 & M2.4 & N18E88 &   1512 &    0.05884 &   17.53 &    0.0006027 &    7.62 &    4.99 &     6.88 &     0.92\\
 235 & 20120818 & 2246 & M1.0 & N18E88 &   1036 &    0.00188 &   12.60 &    0.0000531 &    8.48 &    8.99 &     0.51 &     0.33\\
 238 & 20120906 & 0406 & M1.6 & N04W61 &   2184 &    0.01730 &   22.22 &    0.0000244 &    9.41 &    9.46 &     1.43 &     5.75\\
 241 & 20120930 & 0427 & M1.3 & N12W81 &   2228 &    0.00236 &    9.32 &    0.0009274 &    7.16 &    4.94 &     1.40 &    22.21\\
 245 & 20121020 & 1805 & M9.0 & S12E88 &   2116 &    0.08375 &   10.51 &    0.0036557 &    8.14 &    9.81 &     2.98 &     0.25\\
 246 & 20121021 & 1946 & M1.3 & S13E78 &   2124 &    0.01076 &   19.52 &    0.0004559 &    7.47 &    9.81 &     1.07 &     0.14\\
 248 & 20121023 & 0313 & X1.8 & S13E58 &   1380 &    0.01599 &   26.74 &    0.0562808 &    6.90 &   10.40 &     1.19 &     0.00\\
 251 & 20121112 & 2313 & M2.0 & S25E48 &   2124 &    0.03314 &    8.28 &    0.0002480 &    8.21 &    8.45 &     2.34 &     1.33\\
 253 & 20121113 & 0542 & M2.5 & S26E44 &   1396 &    0.02954 &   21.22 &    0.0003934 &    8.13 &   10.06 &     1.70 &     1.22\\
 255 & 20121114 & 0359 & M1.1 & S23E27 &   1352 &    0.03191 &    6.93 &    0.0014156 &    3.44 &    5.17 &     4.81 &     3.19\\
 257 & 20121120 & 1921 & M1.6 & N10E19 &    372 &    0.04471 &    7.90 &    0.0007343 &    4.91 &    8.61 &     2.65 &     0.18\\
 258 & 20121121 & 0645 & M1.4 & N10E12 &    932 &    0.02454 &    9.04 &    0.0008045 &    6.36 &   11.93 &     1.20 &     0.08\\
 261 & 20121127 & 2105 & M1.0 & S13W42 &   1668 &    0.00753 &   14.83 &    0.0001938 &    7.99 &    7.09 &     1.45 &     0.22\\
 262 & 20121128 & 2120 & M2.2 & S12W56 &   3044 &    0.03893 &   19.23 &    0.0007241 &    7.00 &   12.86 &     1.35 &     0.17\\
 264 & 20130111 & 0843 & M1.2 & N05E42 &   1180 &    0.00542 &    7.63 &    0.0003004 &    7.00 &    7.66 &     1.10 &     0.00\\
 266 & 20130113 & 0045 & M1.0 & N18W15 &    764 &    0.01716 &    7.39 &    0.0011418 &    6.06 &    6.18 &     2.70 &    31.98\\
 268 & 20130217 & 1545 & M1.9 & N12E23 &    620 &    0.01225 &    8.89 &    0.0000633 &    8.87 &    4.87 &     3.26 &     2.45\\
 271 & 20130321 & 2142 & M1.6 & N09W88 &   3516 &    0.03346 &   12.18 &    0.0000383 &    8.23 &   12.28 &     1.34 &     0.12\\
 273 & 20130411 & 0655 & M6.5 & N11E13 &   1076 &    0.04168 &   11.42 &    0.0018528 &    5.27 &   25.55 &     0.50 &     0.91\\
 274 & 20130412 & 1952 & M3.3 & N21W47 &   2012 &    0.02328 &   18.80 &    0.0013568 &    7.29 &   13.87 &     0.93 &     0.18\\
 276 & 20130502 & 0458 & M1.1 & N10W19 &   2380 &    0.00017 &   19.42 &    0.0007521 &    4.69 &    8.24 &     0.17 &     0.00\\
 277 & 20130503 & 1639 & M1.3 & N11W38 &   2872 &    0.00010 &   18.37 &    0.0009633 &    5.20 &    3.04 &     0.59 &     0.37\\
 278 & 20130503 & 1724 & M5.7 & N15E83 &   1316 &    0.03689 &   22.67 &    0.0033001 &    6.85 &   13.27 &     1.26 &     0.07\\
 283 & 20130512 & 2237 & M1.2 & N10E89 &   1872 &    0.00186 &   20.36 &    0.0014919 &    6.05 &   11.68 &     0.34 &     0.15\\
 284 & 20130513 & 0153 & X1.7 & N11E89 &   2496 &    0.10615 &   12.49 &    0.0132431 &    7.65 &   16.33 &     1.56 &     0.11\\
 285 & 20130513 & 1157 & M1.3 & N10E89 &   1048 &    0.00403 &   23.26 &    0.0014927 &    6.72 &    3.52 &     3.04 &     0.00\\
 288 & 20130515 & 0125 & X1.2 & N10E68 &   3524 &    0.09999 &   11.15 &    0.0031250 &    8.06 &   22.63 &     0.93 &     0.83\\
 289 & 20130516 & 2136 & M1.3 & N11E40 &   1280 &    0.00133 &   20.44 &    0.0000784 &    8.12 &    7.27 &     0.59 &     0.15\\
 291 & 20130520 & 0516 & M1.7 & N09E89 &   1380 &    0.01296 &   12.50 &    0.0000855 &    8.06 &    8.08 &     1.57 &     1.98\\
 292 & 20130522 & 1308 & M5.0 & N14W87 &   3248 &    0.04485 &   11.64 &    0.0011678 &    4.63 &   20.27 &     0.73 &     0.18\\
 293 & 20130531 & 1952 & M1.0 & N12E42 &   1060 &    0.00112 &   11.25 &    0.0000235 &    8.27 &    9.35 &     0.37 &     0.06\\
 297 & 20130623 & 2048 & M2.9 & S18E63 &   1132 &    0.02889 &    6.25 &    0.0007958 &    7.29 &    5.01 &     4.79 &     0.06\\
 298 & 20130703 & 0700 & M1.5 & S14E82 &   1548 &    0.01205 &   22.27 &    0.0000406 &    8.91 &    9.38 &     1.21 &     0.13\\
 299 & 20130812 & 1021 & M1.5 & S21E17 &   1536 &    0.00450 &   12.44 &    0.0000636 &    8.64 &   11.58 &     0.54 &     1.24\\
 303 & 20131011 & 0701 & M1.5 & N21E87 &   1124 &    0.01884 &   17.17 &    0.0002881 &    5.13 &    3.48 &     6.69 &     0.64\\
 304 & 20131013 & 0012 & M1.7 & S22E17 &   1416 &    0.67760 &   11.05 &    0.0001016 &    6.80 &    9.52 &     8.86 &     2.12\\
 307 & 20131017 & 1509 & M1.2 & S09W63 &   1696 &    0.00352 &   11.69 &    0.0000092 &    9.04 &   10.42 &     0.56 &     0.07\\
 308 & 20131022 & 0014 & M1.0 & N08E20 &   1068 &    0.00014 &   21.27 &    0.0003649 &    6.90 &    8.32 &     0.15 &     0.00\\
 311 & 20131023 & 2041 & M2.7 & N08W06 &   3368 &    0.01733 &   18.16 &    0.0008089 &    6.76 &    9.50 &     1.42 &     3.60\\
 312 & 20131023 & 2333 & M1.4 & N09W08 &   2000 &    0.01171 &   15.02 &    0.0001602 &    5.42 &    6.49 &     2.07 &     0.24\\
 313 & 20131023 & 2358 & M3.1 & N09W09 &    452 &    0.00031 &   21.46 &    0.0003714 &    7.34 &    8.84 &     0.21 &     0.00\\
 317 & 20131025 & 0248 & M2.9 & S07E76 &   3164 &    0.03163 &   18.68 &    0.0004501 &    7.18 &   12.85 &     1.22 &     0.94\\
 318 & 20131025 & 0753 & X1.7 & S08E73 &    676 &    0.04461 &   33.74 &    0.0298859 &    7.58 &   11.36 &     1.74 &     0.17\\
 320 & 20131025 & 1451 & X2.1 & S06E69 &   3568 &    0.10233 &   11.35 &    0.0003450 &   10.39 &   16.98 &     1.45 &     0.54\\
 321 & 20131025 & 1702 & M1.3 & S08E67 &   2052 &    0.01089 &   15.70 &    0.0008598 &    5.99 &    7.14 &     1.73 &     1.22\\
 325 & 20131026 & 0917 & M1.5 & S08E59 &   1060 &    0.00078 &   11.77 &    0.0001197 &    6.67 &    6.48 &     0.54 &     0.08\\
 328 & 20131026 & 1949 & M1.0 & S08E51 &   1940 &    0.00004 &   20.02 &    0.0001241 &    6.61 &    3.87 &     0.25 &     0.00\\
 334 & 20131028 & 1446 & M2.7 & S08E27 &   2600 &    0.00557 &   19.67 &    0.0006088 &    7.57 &   23.10 &     0.21 &     0.07\\
 336 & 20131028 & 2048 & M1.5 & N07W83 &   1748 &    0.00481 &    8.04 &    0.0002001 &    7.99 &    6.48 &     1.33 &     1.22\\
 340 & 20131102 & 2213 & M1.6 & S12W12 &    768 &    0.00239 &    8.99 &    0.0002536 &    7.73 &    5.47 &     1.21 &     0.19\\
 343 & 20131105 & 1808 & M1.0 & S12E47 &   1124 &    0.00159 &    6.43 &    0.0001669 &    7.77 &    4.57 &     1.29 &     5.16\\
 345 & 20131106 & 1339 & M3.8 & S09E35 &   1936 &    0.00399 &    9.70 &    0.0031537 &    6.63 &    7.92 &     0.90 &     0.15\\
 347 & 20131107 & 0334 & M2.3 & S08E26 &   1436 &    0.02208 &    3.12 &    0.0064472 &    5.08 &   12.92 &     1.01 &     0.04\\
 351 & 20131110 & 0508 & X1.1 & S11W17 &   3284 &    0.04878 &   21.66 &    0.0079130 &    7.69 &   22.03 &     0.68 &     0.20\\
 352 & 20131111 & 1101 & M2.4 & S17E74 &   3068 &    0.00399 &   19.31 &    0.0002777 &    7.71 &   10.35 &     0.60 &     0.10\\
 353 & 20131113 & 1457 & M1.4 & S20E46 &   1400 &    0.00130 &   20.16 &    0.0001988 &    7.48 &   14.63 &     0.20 &     0.07\\
 354 & 20131115 & 0220 & M1.0 & N07E53 &   1252 &    0.00109 &   20.12 &    0.0001849 &    7.62 &    9.28 &     0.37 &     0.03\\
 357 & 20131117 & 0506 & M1.0 & S19W41 &   1208 &    0.00089 &    6.36 &    0.0002105 &    7.49 &    2.98 &     1.84 &     0.46\\
 359 & 20131121 & 1052 & M1.2 & S14W89 &   1248 &    0.02074 &   16.89 &    0.0004123 &    4.71 &    4.55 &     4.69 &     2.51\\
 360 & 20131123 & 0220 & M1.1 & N13W58 &   2584 &    0.00110 &   17.26 &    0.0000888 &    7.99 &    5.71 &     0.77 &     0.25\\
 363 & 20131219 & 2306 & M3.5 & S16E89 &   2304 &    0.01275 &   21.85 &    0.0004127 &    8.06 &   15.14 &     0.61 &     0.04\\
 364 & 20131220 & 1135 & M1.6 & S16E78 &   4272 &    0.00332 &   15.97 &    0.0001171 &    6.64 &    7.27 &     0.93 &     0.58\\
 365 & 20131222 & 0805 & M1.9 & S17W51 &   1788 &    0.00701 &   18.95 &    0.0003829 &    7.68 &    5.42 &     2.10 &     0.25\\
 366 & 20131222 & 0833 & M1.1 & S17W52 &   1956 &    0.00852 &   15.06 &    0.0004831 &    4.45 &    6.18 &     1.90 &     0.28\\
 367 & 20131222 & 1424 & M1.6 & S16E44 &   2532 &    0.03249 &   11.35 &    0.0004446 &    6.57 &    9.85 &     1.84 &     0.06\\
 368 & 20131222 & 1506 & M3.3 & S17W55 &   1328 &    0.00742 &   21.78 &    0.0003082 &    7.26 &   13.71 &     0.54 &     0.37\\
 377 & 20140103 & 1241 & M1.0 & S04E52 &   1000 &    0.02158 &    7.58 &    0.0004308 &    4.84 &    3.83 &     6.20 &     2.66\\
 382 & 20140107 & 0349 & M1.0 & N07E07 &   1432 &    0.00661 &    7.25 &    0.0007455 &    6.14 &    4.20 &     2.99 &     0.39\\
 385 & 20140108 & 0339 & M3.6 & N11W88 &   2016 &    0.01548 &   19.24 &    0.0020340 &    6.89 &    3.83 &     5.25 &     0.52\\
 386 & 20140113 & 2148 & M1.3 & S08W75 &    660 &    0.00086 &    7.92 &    0.0019242 &    6.65 &    2.97 &     1.81 &     0.39\\
 387 & 20140127 & 0105 & M1.0 & S16E88 &   2860 &    0.00172 &   16.77 &    0.0002649 &    4.60 &   11.25 &     0.35 &     0.03\\
 389 & 20140127 & 2205 & M4.9 & S14E88 &   1880 &    0.00078 &   24.20 &    0.0041016 &    6.96 &    4.85 &     0.83 &    61.62\\
 393 & 20140128 & 1233 & M1.3 & S15E79 &   1708 &    0.00363 &    5.90 &    0.0000335 &    8.80 &    4.85 &     1.78 &     0.00\\
 395 & 20140128 & 2204 & M2.6 & S14E74 &   1112 &    0.00399 &    7.01 &    0.0021642 &    6.82 &    5.69 &     1.47 &     0.79\\
\enddata
\end{deluxetable}

\clearpage
%%%%%%%%%%%%%%%%%%%%%%%%%%%%%% TABLE 3 %%%%%%%%%%%%%%%%%%%%%%%%%%%%%%%%%%

\begin{deluxetable}{rrrrrrrrr}
\tabletypesize{\normalsize}
\tabletypesize{\footnotesize}
\tablecaption{Low-energy cutoff energies and total nonthermal energies
calculated for 4 models (en, wt, tof, co), derived from the observables
of 143 M and X-class flare events, tabulated in Table 2.}
\tablewidth{0pt}
\tablehead{
\colhead{ID}&
\colhead{Cutoff}&
\colhead{Cutoff}&
\colhead{Cutoff}&
\colhead{Cutoff}&
\colhead{Nonthermal}&
\colhead{Nonthermal}&
\colhead{Nonthermal}&
\colhead{Nonthermal}\\
\colhead{}&
\colhead{energy}&
\colhead{energy}&
\colhead{energy}&
\colhead{energy}&
\colhead{energy}&
\colhead{energy}&
\colhead{energy}&
\colhead{energy}\\
\colhead{}&
\colhead{$\varepsilon_{en}$}&
\colhead{$\varepsilon_{wt}$}&
\colhead{$\varepsilon_{tof}$}&
\colhead{$\varepsilon_{co}$}&
\colhead{$E_{en}$}&
\colhead{$E_{wt}$}&
\colhead{$E_{tof}$}&
\colhead{$E_{co}$}\\
\colhead{}&
\colhead{(keV)}&
\colhead{(keV)}&
\colhead{(keV)}&
\colhead{(keV)}&
\colhead{(10$^{30}$ erg)}&
\colhead{(10$^{30}$ erg)}&
\colhead{(10$^{30}$ erg)}&
\colhead{(10$^{30}$ erg)}\\}
\startdata
   1 &     1.00 &     4.80 &     6.70 &    15.00 &       0.7778 &       0.0042 &       0.0015 &       0.0001\\
   2 &     3.20 &     6.60 &     1.70 &    19.00 &       0.0305 &       0.0017 &       0.4079 &       0.0000\\
   3 &     0.70 &     5.30 &     1.90 &    30.00 &      21.5602 &       0.0497 &       1.2021 &       0.0002\\
   4 &    11.50 &    14.20 &    11.20 &    21.00 &       2.2251 &       0.4854 &       2.6257 &       0.0307\\
  10 &    16.20 &    13.40 &     9.00 &    30.00 &       1.4205 &       4.5095 &      49.6561 &       0.0333\\
  12 &    12.80 &    14.80 &    15.10 &    27.00 &      46.3275 &      19.9327 &      17.1499 &       0.5138\\
  13 &     8.70 &    12.60 &     5.90 &    21.00 &       2.4624 &       0.2781 &      23.6050 &       0.0143\\
  15 &     9.20 &     8.90 &     9.90 &    22.00 &       5.1154 &       6.0449 &       3.1357 &       0.0165\\
  16 &    14.90 &     6.90 &    15.50 &    27.00 &       2.5982 &     312.6780 &       2.0072 &       0.0636\\
  18 &    10.50 &     9.60 &    10.90 &    24.00 &       0.8962 &       1.6026 &       0.7330 &       0.0057\\
  19 &     1.90 &     3.20 &     9.70 &    15.00 &       0.3648 &       0.0681 &       0.0017 &       0.0004\\
  20 &    10.70 &     5.80 &    10.40 &    22.00 &       2.0033 &     143.3275 &       2.4943 &       0.0132\\
  21 &     9.90 &     9.60 &     7.50 &    15.00 &       5.4652 &       7.0165 &      52.4010 &       0.1755\\
  22 &     7.60 &     5.50 &     5.30 &    22.00 &       5.5665 &      35.2264 &      41.2397 &       0.0164\\
  23 &     9.80 &     5.80 &     6.40 &    20.00 &       3.4690 &      90.9612 &      49.9326 &       0.0367\\
  28 &     5.40 &     5.60 &     4.50 &    15.00 &       6.6379 &       5.5572 &      14.3126 &       0.0957\\
  29 &    12.90 &     6.00 &     5.20 &    26.00 &       0.1107 &       4.2193 &       8.2575 &       0.0039\\
  30 &     9.20 &     2.70 &     9.60 &    22.00 &       0.9000 &     895.5553 &       0.7023 &       0.0063\\
  31 &     7.90 &     4.50 &    38.30 &    30.00 &      16.7310 &     975.3401 &       0.0002 &       0.0012\\
  33 &    29.20 &    16.20 &     6.60 &    21.00 &       0.0294 &       1.3916 &     495.0242 &       0.2530\\
  34 &     6.10 &     7.70 &     5.80 &    15.00 &     416.3836 &      68.0520 &     624.6119 &       0.2839\\
  37 &    12.80 &     8.40 &     9.30 &    28.00 &       5.1082 &      41.4942 &      25.9488 &       0.0969\\
  38 &    10.00 &     5.70 &    11.80 &    20.00 &       0.6242 &      26.0680 &       0.2079 &       0.0063\\
  40 &    11.30 &     7.40 &    18.20 &    30.00 &       3.8600 &      47.7786 &       0.2403 &       0.0126\\
  41 &     4.10 &     4.70 &    13.20 &    15.00 &       1.1429 &       0.6713 &       0.0102 &       0.0060\\
  46 &     8.80 &     8.20 &     7.90 &    20.00 &       7.1019 &      10.9290 &      13.8560 &       0.0568\\
  48 &    96.70 &     8.20 &     8.80 &    16.00 &       0.0000 &       5.1251 &       3.4567 &       0.0888\\
  50 &     2.00 &     3.10 &    35.60 &    12.00 &       2.7037 &       0.6692 &       0.0005 &       0.0127\\
  51 &     6.60 &     7.90 &     6.50 &    12.00 &      14.6050 &       4.8811 &      16.0264 &       0.3311\\
  52 &     8.10 &     9.80 &     6.40 &    15.00 &      10.3822 &       2.4438 &      71.7504 &       0.0802\\
  53 &    14.20 &    13.00 &    21.20 &    15.00 &       5.3108 &       9.4608 &       0.3361 &       3.6127\\
  55 &    54.60 &     9.60 &     7.70 &    19.00 &       0.0001 &       5.8205 &      24.7286 &       0.0624\\
  61 &    14.50 &    18.60 &    13.90 &    28.00 &     106.4098 &      21.3920 &     137.7906 &       1.5931\\
  63 &    10.40 &    11.30 &     5.10 &    18.00 &       1.3532 &       0.7893 &     198.1350 &       0.0300\\
  64 &     8.10 &    11.30 &     7.10 &    30.00 &       2.6708 &       0.2223 &       6.7030 &       0.0002\\
  65 &    12.00 &     8.10 &     9.30 &    30.00 &      22.9111 &     404.4191 &     152.0502 &       0.0251\\
  68 &    10.50 &    16.70 &    14.30 &    18.00 &     101.2172 &       3.2939 &      10.3612 &       1.9095\\
  69 &    10.50 &     6.70 &     9.60 &    16.00 &       9.9118 &     168.5142 &      17.3350 &       0.7104\\
  70 &     7.60 &    10.80 &     6.60 &    17.00 &      19.8497 &       1.0888 &      68.6569 &       0.0234\\
  71 &    21.20 &    16.10 &     4.80 &    30.00 &       0.0204 &       0.1367 &     570.6394 &       0.0019\\
  77 &     9.00 &     7.50 &     6.00 &    15.00 &       3.5533 &      11.7135 &      50.8674 &       0.1313\\
  81 &    13.40 &     6.90 &    11.00 &    23.00 &       0.7741 &      59.5029 &       2.8106 &       0.0225\\
  83 &     6.30 &     6.60 &     8.00 &    14.00 &      98.3267 &      65.8536 &      19.0484 &       0.4112\\
  84 &     8.60 &     5.70 &    14.90 &    29.00 &       6.2511 &      51.7382 &       0.4210 &       0.0151\\
  86 &     8.80 &    10.20 &     5.80 &    28.00 &       1.7223 &       0.6667 &      31.8594 &       0.0007\\
  91 &     7.00 &     7.20 &     8.30 &    12.00 &     135.1171 &     107.9637 &      37.2981 &       2.0702\\
  98 &    15.90 &     9.70 &     8.70 &    20.00 &       1.0634 &      39.2940 &      82.8666 &       0.2042\\
 100 &    14.90 &    14.90 &     8.40 &    20.00 &       0.6826 &       0.7121 &      38.6946 &       0.0894\\
 101 &    11.20 &     4.90 &    11.70 &    15.00 &       0.0898 &      13.9602 &       0.0692 &       0.0157\\
 103 &    12.80 &    13.70 &    10.50 &    28.00 &       0.9179 &       0.5842 &       2.9818 &       0.0077\\
 111 &    10.30 &    13.20 &     4.90 &    22.00 &       1.4515 &       0.2686 &     201.8147 &       0.0086\\
 116 &    23.70 &     7.10 &     5.80 &    14.00 &       0.0077 &      47.6822 &     205.9947 &       0.3427\\
 120 &     7.70 &     5.80 &     9.50 &    15.00 &       6.5091 &      44.8960 &       1.4804 &       0.0639\\
 122 &    10.00 &    12.00 &     7.40 &    27.00 &       2.4833 &       0.6709 &      21.1648 &       0.0022\\
 123 &     9.90 &    11.50 &     5.80 &    28.00 &       4.1542 &       1.4337 &     172.7290 &       0.0031\\
 125 &     8.20 &    14.50 &     7.30 &    26.00 &       6.0413 &       0.1924 &      11.8059 &       0.0056\\
 126 &     8.40 &    11.00 &     6.40 &    22.00 &      13.1490 &       1.9445 &      99.5300 &       0.0130\\
 157 &    36.50 &     9.50 &     7.40 &    14.00 &       0.0001 &       3.0211 &      21.6221 &       0.1610\\
 158 &     5.30 &     7.60 &     7.40 &    15.00 &       1.3581 &       0.2412 &       0.2706 &       0.0095\\
 159 &     7.30 &    11.50 &    12.30 &    17.00 &       7.0812 &       0.5296 &       0.3739 &       0.0578\\
 160 &     5.90 &     2.10 &    12.70 &    17.00 &       3.0419 &     473.1884 &       0.0663 &       0.0157\\
 167 &     0.40 &     4.60 &     8.60 &    21.00 &      30.2518 &       0.0892 &       0.0196 &       0.0022\\
 168 &     9.70 &     8.30 &     6.50 &    15.00 &       1.7269 &       4.1018 &      15.3409 &       0.1610\\
 169 &     6.60 &     8.60 &    11.00 &    15.00 &     110.4820 &      16.9623 &       3.0718 &       0.3498\\
 170 &     0.70 &     1.60 &    13.00 &    14.00 &      19.2109 &       1.2400 &       0.0019 &       0.0015\\
 173 &    13.90 &     6.10 &    10.20 &    20.00 &       0.4422 &     140.1815 &       3.7789 &       0.0338\\
 176 &     2.90 &     5.10 &    12.80 &    15.00 &       0.5684 &       0.0932 &       0.0047 &       0.0028\\
 178 &     9.10 &     7.90 &    12.50 &    19.00 &       1.6227 &       3.9857 &       0.1907 &       0.0118\\
 182 &     7.30 &     9.30 &     6.60 &    30.00 &      10.8205 &       2.1517 &      21.4655 &       0.0008\\
 187 &    11.40 &     9.40 &    11.40 &    22.00 &      10.0191 &      33.7559 &      10.0593 &       0.1921\\
 189 &     0.60 &     4.50 &    12.30 &    16.00 &       0.6845 &       0.0060 &       0.0006 &       0.0003\\
 190 &     0.10 &     3.90 &     8.00 &    14.00 &       4.6077 &       0.0036 &       0.0010 &       0.0004\\
 195 &    12.80 &     6.80 &    12.40 &    24.00 &       4.9962 &     219.9091 &       5.9444 &       0.1149\\
 196 &     9.30 &     8.60 &     6.80 &    22.00 &       3.5504 &       6.0006 &      25.6911 &       0.0142\\
 199 &     6.30 &     7.50 &     9.70 &    18.00 &       5.3694 &       2.2979 &       0.6250 &       0.0267\\
 200 &     1.00 &     8.20 &     1.80 &    15.00 &      40.7531 &       0.0202 &       5.0761 &       0.0024\\
 203 &    14.50 &    10.00 &    11.40 &    21.00 &       0.4537 &       6.5307 &       2.5106 &       0.0320\\
 205 &     5.70 &     8.30 &    10.80 &    22.00 &       5.8572 &       0.9418 &       0.2667 &       0.0089\\
 208 &     8.60 &    10.00 &     7.60 &    24.00 &       6.2984 &       2.6349 &      14.0258 &       0.0131\\
 210 &    11.70 &    12.80 &     8.10 &    24.00 &       0.4536 &       0.2598 &       4.1385 &       0.0062\\
 211 &     5.90 &    10.30 &     6.90 &    30.00 &       2.0414 &       0.0798 &       0.8681 &       0.0001\\
 212 &     4.90 &    15.30 &     9.90 &    24.00 &     505.2834 &       0.4216 &       6.1786 &       0.0257\\
 214 &    19.60 &    12.80 &     6.10 &    30.00 &       0.0259 &       0.5373 &     105.4782 &       0.0012\\
 215 &     5.70 &     8.90 &    11.40 &    17.00 &       8.9478 &       0.8496 &       0.2419 &       0.0290\\
 219 &     9.30 &    13.00 &     6.00 &    30.00 &       7.8497 &       0.9755 &     110.5648 &       0.0055\\
 223 &    14.80 &     7.50 &    14.20 &    16.00 &       0.7552 &      30.5329 &       0.9672 &       0.5032\\
 228 &     5.40 &     3.40 &    16.20 &    15.00 &       0.7626 &       5.0530 &       0.0073 &       0.0101\\
 230 &    11.90 &    13.00 &    16.40 &    18.00 &       0.7520 &       0.4090 &       0.0897 &       0.0480\\
 235 &     9.40 &    10.30 &     6.00 &    30.00 &       2.7705 &       1.4368 &      84.1892 &       0.0005\\
 238 &     8.30 &    19.90 &    10.30 &    17.00 &      26.7858 &       0.0172 &       4.5569 &       0.0661\\
 241 &     7.60 &     6.60 &     7.30 &    16.00 &       9.4226 &      23.5095 &      11.6637 &       0.0967\\
 245 &    15.90 &     8.30 &    15.10 &    20.00 &       2.5768 &     276.5678 &       3.7736 &       0.5087\\
 246 &    11.40 &    14.30 &     9.00 &    22.00 &       1.3530 &       0.3115 &       5.9178 &       0.0188\\
 248 &   194.30 &    18.20 &     9.80 &    30.00 &       0.0000 &       0.9979 &      37.5929 &       0.0524\\
 251 &    10.00 &     6.60 &    12.40 &    22.00 &       5.8906 &     121.1315 &       1.2253 &       0.0201\\
 253 &     9.40 &    16.70 &    11.60 &    20.00 &      10.7416 &       0.1714 &       2.3528 &       0.0474\\
 255 &     0.80 &     2.60 &    13.90 &    14.00 &       0.1509 &       0.0085 &       0.0001 &       0.0001\\
 257 &     3.20 &     4.00 &    13.30 &    16.00 &       0.9894 &       0.3824 &       0.0035 &       0.0017\\
 258 &     7.70 &     5.70 &    10.60 &    17.00 &       3.0739 &      14.6470 &       0.5495 &       0.0434\\
 261 &    11.90 &    11.50 &     9.00 &    24.00 &       0.6241 &       0.8022 &       4.5538 &       0.0047\\
 262 &     9.70 &    13.30 &    11.70 &    17.00 &       2.8245 &       0.4292 &       0.9263 &       0.0964\\
 264 &    32.10 &     5.30 &     8.10 &    18.00 &       0.0016 &      80.4219 &       6.0352 &       0.0504\\
 266 &     3.60 &     4.50 &    11.40 &    15.00 &      24.2342 &       8.1539 &       0.0734 &       0.0184\\
 268 &     9.70 &     7.60 &    11.10 &    30.00 &       2.4258 &      17.6247 &       0.8419 &       0.0003\\
 271 &     9.90 &     9.70 &    11.40 &    14.00 &       3.0094 &       3.6091 &       1.1474 &       0.2522\\
 273 &     2.30 &     6.20 &    10.00 &    12.00 &      83.7943 &       1.2304 &       0.1574 &       0.0719\\
 274 &    10.40 &    13.40 &    10.10 &    16.00 &      13.3241 &       2.7140 &      16.7779 &       0.9044\\
 276 &    52.20 &     9.50 &     3.30 &    16.00 &       0.0001 &       0.0280 &       1.3581 &       0.0041\\
 277 &     9.50 &     9.80 &     3.70 &    21.00 &       0.1121 &       0.0971 &       5.6071 &       0.0040\\
 278 &    11.60 &    15.30 &    11.40 &    20.00 &       3.1159 &       0.5936 &       3.3443 &       0.1256\\
 283 &     7.80 &    12.40 &     5.60 &    29.00 &       1.1321 &       0.1071 &       5.9789 &       0.0014\\
 284 &    15.80 &     9.30 &    14.10 &    25.00 &      14.9141 &     511.5092 &      32.2520 &       0.7191\\
 285 &    77.60 &    15.50 &     9.10 &    28.00 &       0.0000 &       0.1121 &       2.2906 &       0.0038\\
 288 &    10.30 &     8.70 &    12.80 &    18.00 &      70.1435 &     233.2775 &      15.2921 &       1.3882\\
 289 &    11.10 &    16.10 &     5.80 &    30.00 &       1.2008 &       0.0851 &     122.8816 &       0.0010\\
 291 &     7.80 &     9.80 &     9.90 &    15.00 &      10.7964 &       2.1221 &       1.8687 &       0.1025\\
 292 &     2.60 &     5.60 &    10.80 &    14.00 &       8.6450 &       0.5070 &       0.0486 &       0.0188\\
 293 &     9.80 &     9.00 &     5.20 &    24.00 &       0.6634 &       1.2131 &      65.5062 &       0.0010\\
 297 &    15.80 &     4.50 &    13.70 &    15.00 &       0.1384 &     394.0855 &       0.3433 &       0.1936\\
 298 &    11.60 &    19.00 &     9.40 &    16.00 &       1.5009 &       0.0292 &       7.6938 &       0.1153\\
 299 &     8.20 &    10.30 &     7.00 &    15.00 &      15.9317 &       2.7585 &      56.0776 &       0.1611\\
 303 &     5.00 &     9.10 &    13.50 &    17.00 &       0.2404 &       0.0207 &       0.0040 &       0.0016\\
 304 &     4.80 &     7.40 &    25.70 &    30.00 &       8.2187 &       0.6604 &       0.0005 &       0.0002\\
 307 &    10.50 &    10.10 &     6.70 &    23.00 &       2.4395 &       3.3935 &      89.3478 &       0.0046\\
 308 &    39.30 &    14.50 &     3.20 &    30.00 &       0.0002 &       0.0829 &     654.5867 &       0.0011\\
 311 &     6.80 &    12.20 &    10.30 &    21.00 &      21.6912 &       0.7447 &       1.9624 &       0.0318\\
 312 &     5.00 &     8.30 &    10.20 &    21.00 &       0.1779 &       0.0181 &       0.0072 &       0.0003\\
 313 &    20.70 &    15.40 &     3.80 &    30.00 &       0.0136 &       0.0883 &     606.1953 &       0.0013\\
 317 &     7.60 &    13.20 &    11.10 &    20.00 &      20.5291 &       0.6779 &       1.9889 &       0.0513\\
 318 &    16.00 &    25.00 &    12.40 &    30.00 &      10.2022 &       0.5564 &      54.3011 &       0.1656\\
 320 &    14.60 &    11.10 &    13.80 &    30.00 &      14.5203 &     178.0020 &      23.1600 &       0.0163\\
 321 &     6.30 &     9.50 &     9.80 &    16.00 &       1.8441 &       0.2419 &       0.2003 &       0.0176\\
 325 &     8.70 &     7.80 &     5.20 &    20.00 &       0.5292 &       1.0053 &       9.8712 &       0.0048\\
 328 &    26.10 &    13.10 &     2.70 &    30.00 &       0.0003 &       0.0150 &      97.9885 &       0.0001\\
 334 &     9.70 &    14.50 &     6.20 &    21.00 &       4.4813 &       0.3091 &      84.4325 &       0.0276\\
 336 &    10.00 &     6.20 &     8.20 &    17.00 &       2.2796 &      64.5426 &       9.3836 &       0.0578\\
 340 &    11.80 &     6.80 &     7.20 &    17.00 &       0.7679 &      32.6484 &      21.8765 &       0.0662\\
 343 &     8.30 &     4.90 &     6.80 &    21.00 &       3.3767 &     129.3156 &      13.4555 &       0.0064\\
 345 &    12.80 &     6.40 &     7.40 &    21.00 &       1.1343 &      57.4849 &      24.0198 &       0.0702\\
 347 &     7.40 &     1.60 &    10.10 &    14.00 &       0.7488 &     361.8527 &       0.2138 &       0.0565\\
 351 &    12.80 &    16.20 &    10.80 &    20.00 &      10.7767 &       2.1711 &      33.3844 &       0.5354\\
 352 &    12.10 &    14.50 &     7.00 &    23.00 &       2.4294 &       0.7171 &      98.3107 &       0.0324\\
 353 &     9.00 &    14.70 &     4.80 &    27.00 &       4.3452 &       0.1747 &     244.4080 &       0.0035\\
 354 &    12.20 &    15.00 &     5.20 &    26.00 &       0.5077 &       0.1293 &     145.9955 &       0.0033\\
 357 &    13.10 &     4.70 &     6.50 &    15.00 &       0.1348 &     109.7154 &      12.1532 &       0.0556\\
 359 &     2.70 &     8.30 &    12.90 &    15.00 &       2.9134 &       0.0436 &       0.0085 &       0.0049\\
 360 &    12.20 &    13.40 &     5.90 &    15.00 &       1.2240 &       0.6325 &     201.4743 &       0.2834\\
 363 &    12.70 &    17.10 &     8.50 &    21.00 &       2.8401 &       0.3577 &      50.4921 &       0.0826\\
 364 &     7.40 &    10.50 &     7.30 &    17.00 &       1.8018 &       0.2566 &       2.0618 &       0.0171\\
 365 &    13.30 &    14.20 &     9.40 &    22.00 &       0.3626 &       0.2367 &       3.6241 &       0.0126\\
 366 &     3.30 &     7.10 &     9.60 &    15.00 &       0.1169 &       0.0084 &       0.0030 &       0.0006\\
 367 &    10.10 &     7.40 &    11.90 &    14.00 &       0.7798 &       4.3741 &       0.3099 &       0.1260\\
 368 &     7.30 &    15.50 &     7.60 &    22.00 &       7.7331 &       0.0665 &       5.8546 &       0.0074\\
 377 &     3.10 &     3.80 &    13.60 &    16.00 &       0.6630 &       0.3083 &       0.0023 &       0.0013\\
 382 &     9.70 &     4.50 &     9.90 &    17.00 &       0.1497 &       7.9724 &       0.1325 &       0.0082\\
 385 &    14.80 &    13.10 &    12.50 &    16.00 &       0.4465 &       0.9144 &       1.1817 &       0.2800\\
 386 &    13.70 &     5.20 &     6.50 &    15.00 &       0.1604 &      36.7204 &      10.8544 &       0.0947\\
 387 &     3.80 &     8.10 &     5.50 &    15.00 &       0.3818 &       0.0258 &       0.1018 &       0.0028\\
 389 &     7.50 &    16.60 &     5.60 &    30.00 &      39.5556 &       0.3435 &     224.5840 &       0.0101\\
 393 &    25.90 &     5.00 &     8.20 &    30.00 &       0.0023 &     900.0698 &      18.1551 &       0.0007\\
 395 &    10.60 &     4.70 &     8.10 &    18.00 &       3.1501 &     350.3838 &      15.3367 &       0.1457\\
\enddata
\end{deluxetable}

\clearpage
%%%%%%%%%%%%%%%%%%%%%%%%%%%%%% FIGURES %%%%%%%%%%%%%%%%%%%%%%%%%%%%%%%%%%

\begin{figure}
\centerline{\includegraphics[width=1.0\textwidth]{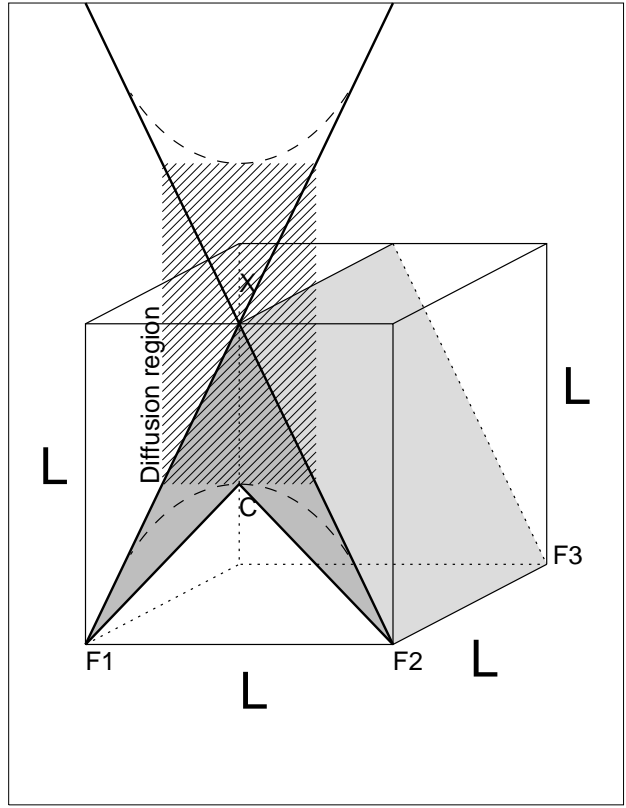}}
\caption{Geometric model of a flare arcade embedded in a cube with length $L$,
width $w=L$, and height $h=L$, with volume $V=L \times w \times h = L^3$.
The footpoints of the loop arcade are at the locations $F1$ and $F2$,
the X-point $X$ at height $h=L$, and the cusp $C$ at height $L/2$.
The magnetic field line through the cusp is approximated with the triangle
$F1-C-F2$ and has the volume $V=L^3 \times q_{geo}$, where the geometric
filling factor of the cube is $q_{geo}=1/4$. The diffusion region of
magnetic reconnection in the X-point is indicated with a shaded area
and has the same filling factor of $q_{geo}=1/4$.}
\end{figure}

\begin{figure}
\centerline{\includegraphics[width=0.7\textwidth]{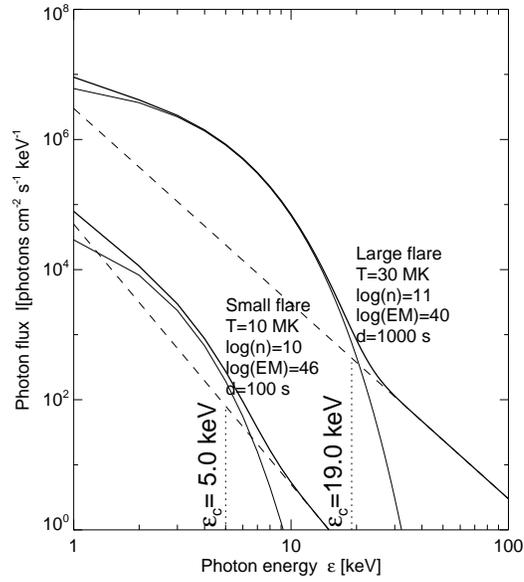}}
\caption{Theoretical hard X-ray spectrum consisting of a
thermal and a nonthermal (power law) component with equal
energy content above the cutoff energy $\varepsilon_c$.
The parameters are chosen for a large flare with $T_e=30$ MK,
$n_e=10^{11}$ cm$^{-3}$, $EM_V=10^{49}$ cm$^{-3}$,
$\gamma=3$, and duration $\tau_{flare}=1000$ s; and
for a small flare with $T_e=10$ MK, $n_e=10^{10}$ cm$^{-3}$,
$EM_V=10^{46}$ cm$^{-3}$, $\gamma=5$, and duration
$\tau_{flare}=100$ s. The x-axis is the photon energy in
units of keV.}
\end{figure}

\begin{figure}
\centerline{\includegraphics[width=0.9\textwidth]{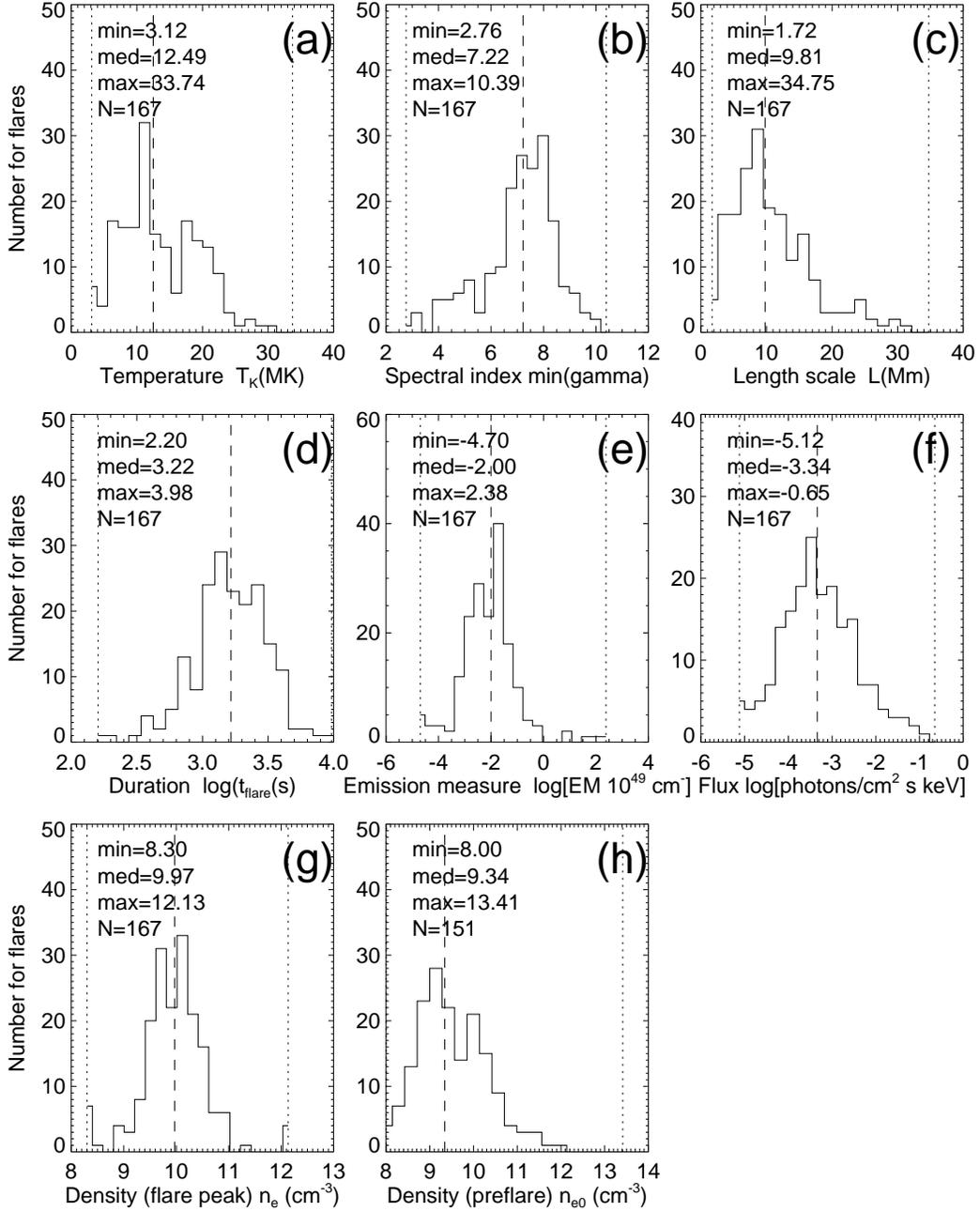}}
\caption{The distributions of measured observables are shown,
which are required in modeling of the low-energy cutoff of
hard X-ray spectra of M and X-class flares. The minimum,
maximum and median values are indicated.}
\end{figure}

\begin{figure}
\centerline{\includegraphics[width=0.9\textwidth]{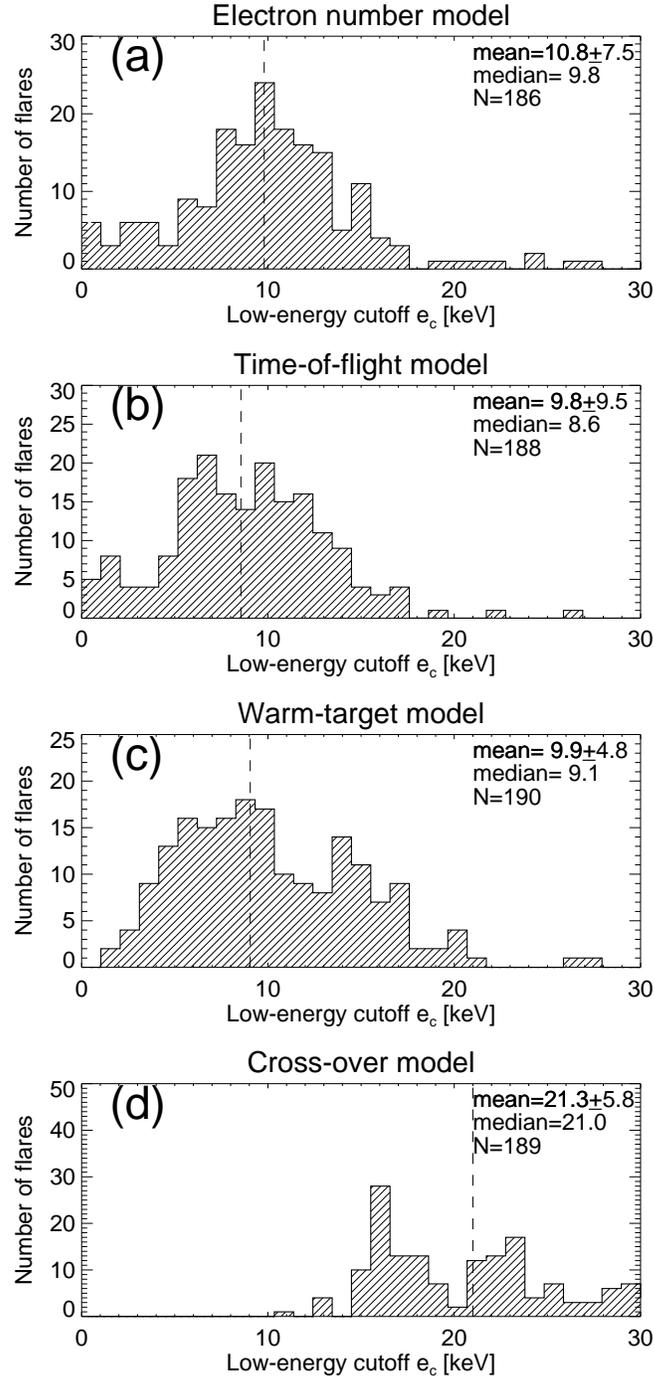}}
\caption{Distribution of low-energy cutoffs in four models:
the electron number model (a), the time-of-flight model (b),
the warm target model (c), and the cross over model (d).
Note that the first three models all yield a low-energy
cutoff energy of $e_c \approx 10$ keV, while the cross-order
model predicts upper limits only, at $e_c \approx 21$ keV.}
\end{figure}

\begin{figure}
\centerline{\includegraphics[width=0.9\textwidth]{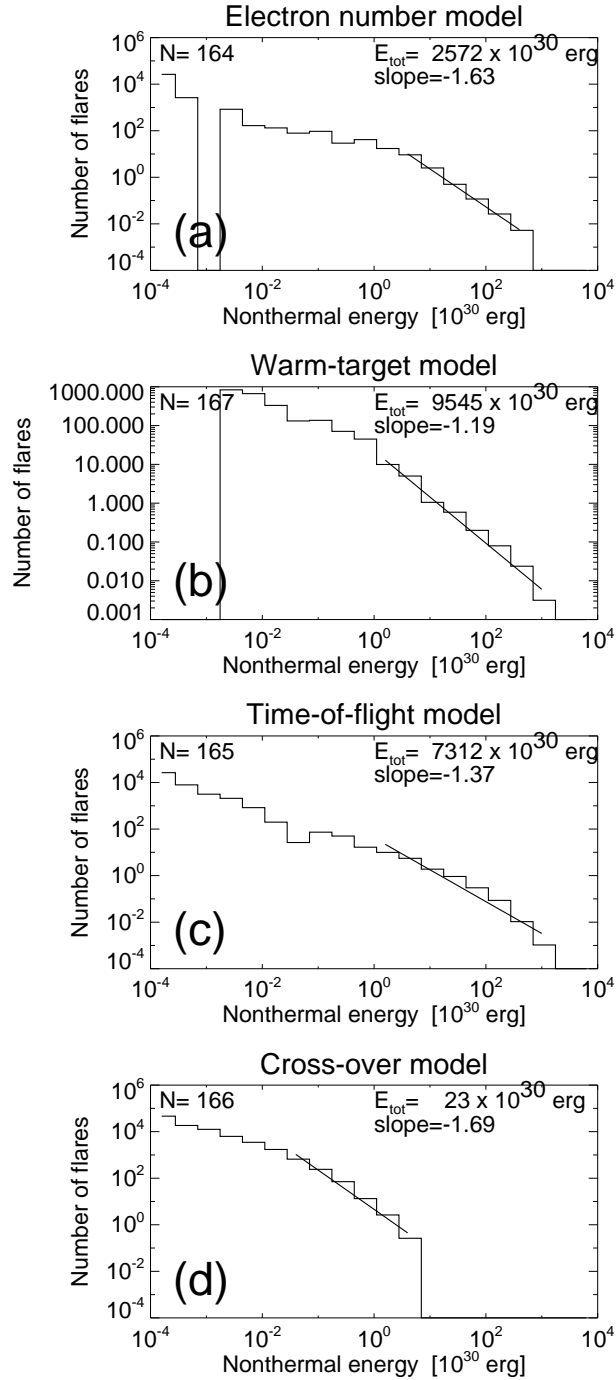}}
\caption{Size distributions of nonthermal energies $E_{nth}$
(histograms with power law fits) and total nonthermal energy $E_{tot}$
contained in each distribution for 4 different models of
the low-energy cutoff $e_c$.}
\end{figure}

\begin{figure}
\centerline{\includegraphics[width=0.9\textwidth]{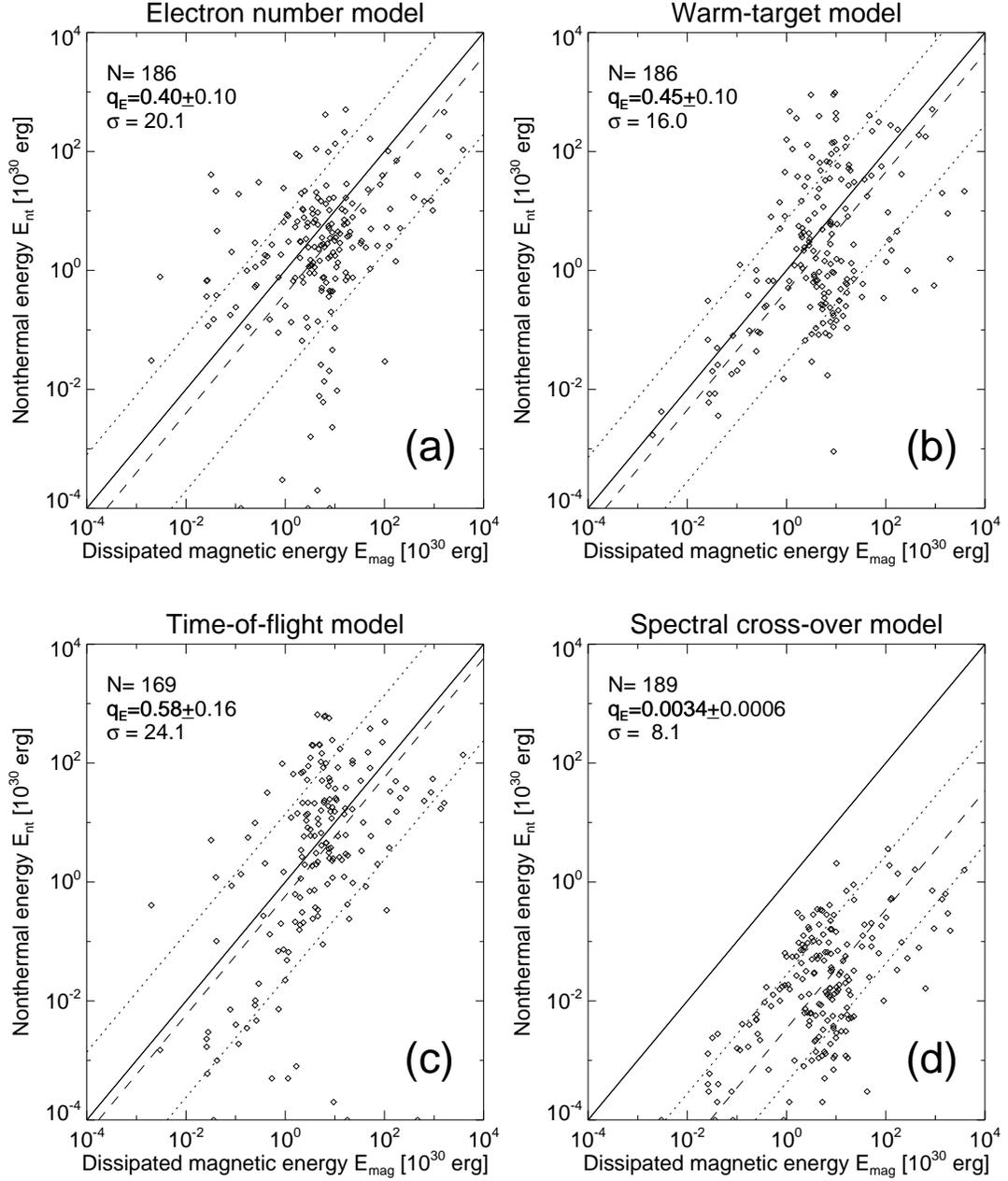}}
\caption{Scatterplots of nonthermal energy in accelerated
electrons ($E_{nth}$) as a function of the dissipated
magnetic energy ($E_{diss}$). Equivalence is rendered with
a solid diagonal line, the logarithmically averaged ratios
$(q_E$) with a dashed line, and the standard deviation
factors ($\sigma$) with dotted lines. }
\end{figure}

\begin{figure}
\centerline{\includegraphics[width=1.0\textwidth]{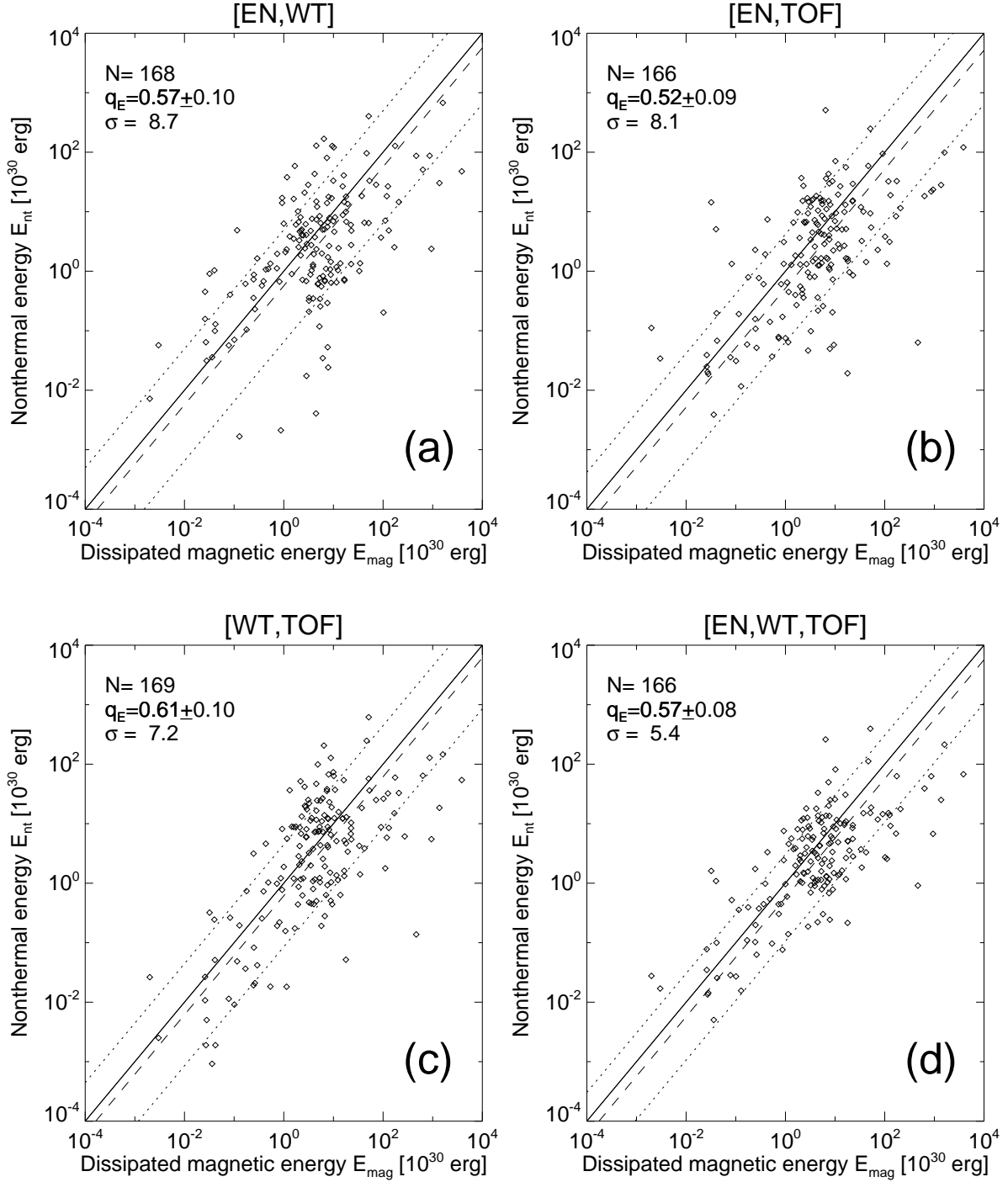}}
\caption{Scatterplots of nonthermal energy in accelerated
electrons $(E_{nth})$ as a function of the dissipated magnetic
energy $(E_{diss})$, averaged from two or three methods:
(a) [EN,WT], (b) [EN, TOF], (c) [WT, TOF],
and (d) [EN, WT, TOF], with similar representation as Fig.~6/
Note that the logarithmically averaged ratios are compatible
with the previous result of $E_{nth}/E_{diss}=0.51\pm0.17$
(Aschwanden et al.~(2017).}
\end{figure}

\begin{figure}
\centerline{\includegraphics[width=1.0\textwidth]{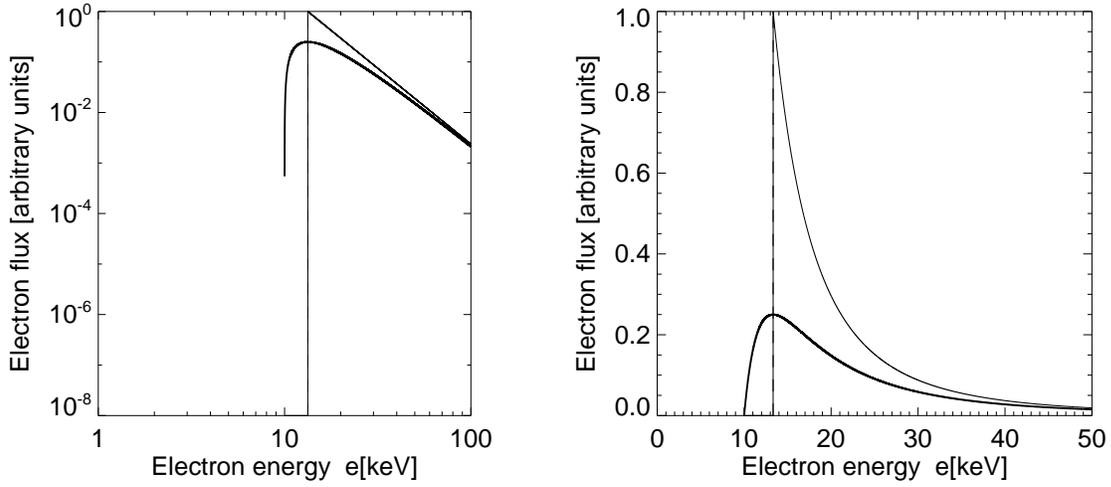}}
\caption{Electron injection spectrum with a
smooth low-energy cutoff at the lower end (thick linestyle),
asymptotically approaching a power law function at the upper end
(as defined in Eq.~A1), rendered on a log-log scale
(left panel), as well as on a linear scale (right panel).
The power law slope of the electron injection spectrum is
$\delta=3$, the minimum energy is $\varepsilon_{min}=10$ keV,
the peak energy is $\varepsilon_{peak}=13.3$ keV,
with ratio $\varepsilon_{peak} / \varepsilon_{min} =(1+1/\delta)=4/3$.}
\end{figure}

\end{document}